\documentclass[review]{elsarticle}

\usepackage{lineno,hyperref}

\modulolinenumbers[5]

\usepackage{rotating}
\usepackage{array}
\usepackage{caption}
\usepackage{ragged2e}
\usepackage[acronym]{glossaries}

\usepackage[nolist]{acronym} 			
\usepackage{booktabs}					
\usepackage{amsmath,amssymb,amsthm} 	
\newcommand{\ra}[1]{\renewcommand{\arraystretch}{#1}}	
\usepackage{subfig}						
\usepackage{tikz}						
\usetikzlibrary{trees,calc}					
\usepackage{comment}					
\usepackage[official]{eurosym}			
\usepackage[at]{easylist}               
\usepackage{natbib}

\usepackage{color,soul}

\usepackage{amsmath}
\usepackage{accents}

\usepackage{lscape}
\usepackage{subfig}
\usepackage{url}

\usepackage{tikz}
\usetikzlibrary{positioning, shapes.geometric}

\makeatletter
\newcounter{manualsubequation}
\renewcommand{\themanualsubequation}{\alph{manualsubequation}}
\newcommand{\startsubequation}{%
  \setcounter{manualsubequation}{0}%
  \refstepcounter{equation}\ltx@label{manualsubeq\theequation}%
  \xdef\labelfor@subeq{manualsubeq\theequation}%
}
\newcommand{\tagsubequation}{%
  \stepcounter{manualsubequation}%
  \tag{\ref{\labelfor@subeq}\themanualsubequation}%
}
\let\subequationlabel\ltx@label
\makeatother

\usepackage{nomencl}
\usepackage{framed}
\usepackage{multicol}
\makenomenclature

\journal{Environmental Research: Energy}

\biboptions{sort&compress} 

\usepackage{anysize}
\marginsize{2.5cm}{2.4cm}{1.5cm}{2cm}

\UseRawInputEncoding

\begin{document}

\begin{frontmatter}

\title{Measuring the Dunkelflaute: How (not) to analyze variable renewable energy shortage}



\author[A1,A2]{Martin Kittel\corref{MK}}
\ead{mkittel@diw.de}

\author[A1]{Wolf-Peter Schill} 
\ead{wschill@diw.de}

\cortext[MK]{Corresponding author}
\address[A1] {DIW Berlin, Department of Energy, Transportation, Environment, Mohrenstra{\ss}e 58, 10117 Berlin, Germany}
\address[A2] {Technical University Berlin, Digital Transformation in Energy Systems, Einsteinufer 25 (TA 8), 10587 Berlin, Germany}




\begin{abstract}

As variable renewable energy sources increasingly gain importance in global energy systems, there is a growing interest in understanding periods of variable renewable energy shortage (``Dunkelflauten''). Defining, quantifying, and comparing such shortage events across different renewable generation technologies and locations presents a surprisingly intricate challenge. Various methodological approaches exist in different bodies of literature, which have been applied to single technologies in specific locations or technology portfolios across multiple regions. We provide an overview of various methods for quantifying variable renewable energy shortage, focusing either on supply from variable renewables or its mismatch with electricity demand. We explain and critically discuss the merits and challenges of different approaches for defining and identifying shortage events and propose further methodological improvements for more accurate shortage determination. Additionally, we elaborate on comparability requirements for multi-technological and multi-regional energy shortage analysis. In doing so, we aim to contribute to unifying disparate methodologies, harmonizing terminologies, and providing guidance for future research.

\vspace{.5cm}


\begin{acronym}
 \acro{CAZ}{Constantly-Above-Zero}
 \acro{CBT}{Constantly-Below-Threshold}
 \acro{FLH}{Full-load hours}
 \acro{FMAZ}{fixed-duration Mean-Above-Zero}
 \acro{FMBT}{fixed-duration Mean-Below-Threshold}
 \acro{MAZ}{Mean-Above-Zero}
 \acro{MBT}{Mean-Below-Threshold}
 \acro{PRL}{positive residual load}
 \acro{PV}{photovoltaics}
 \acro{SPA}{Sequent Peak Algorithm}
 \acro{SPA$^{adj}$}{adjusted Sequent Peak Algorithm}
 \acro{VMAZ}{variable-duration Mean-Above-Zero}
 \acro{VMBT}{variable-duration Mean-Below-Threshold}
 \acro{VRE}{variable renewable energy}
\end{acronym}

\end{abstract}

\begin{keyword}
Variable renewable energy \sep Dunkelflaute \sep Energy shortage \sep Variable renewable energy droughts \sep Positive residual load events \sep Methodological exploration

\end{keyword}

\end{frontmatter}

\section{Introduction}

Wind and solar power are cornerstone technologies for achieving climate-neutral energy systems around the world \cite{lee_ipcc_2023}. Because of their variable generation profiles, their system integration requires an increasing amount of temporal and spatial power sector flexibility. This includes, for example, different types of energy storage, demand-side flexibility, and geographical balancing \cite{denholm_grid_2011,rasmussen_storage_2012,schlachtberger_benefits_2017,roth_geographical_2023,he_sector_2021,kirchem_power_2023}. While various implications of renewable variability have been explored in the literature, attention has recently shifted to the occurrence of prolonged periods with extremely low wind and/or solar availability, also referred to as ``Dunkelflauten''. This applies to both the research and the energy policy discourse in many countries \cite{dawkins_characterising_2020,huneke_f_kalte_2017,dhu_hintergrundpapier_2021,deutscher_bundestag_sicherstellung_2019,wood_go_2021,novacheck_evolving_2021}.

There is a growing interest in the questions as to how frequent and severe such shortage events may become in future energy systems heavily relying on \ac{VRE}, and how to deal with them.  Periods of \ac{VRE} shortage will particularly require the use of long-duration energy storage \cite{dowling_role_2020,sepulveda_design_2021} as the potential for geographical balancing may be limited. This is because of the synoptic scale of these weather events, arising across thousands of kilometers \cite{brown_meteorology_2021}. However, a unique and generally accepted definition of \ac{VRE} shortage events has not yet been established. This is because defining and quantifying such events is a surprisingly intricate task that can be addressed from different angles, requiring a range of methodological choices and parameter assumptions. With this paper, we aim to provide comprehensive guidance for the analysis of variable renewable energy shortage.

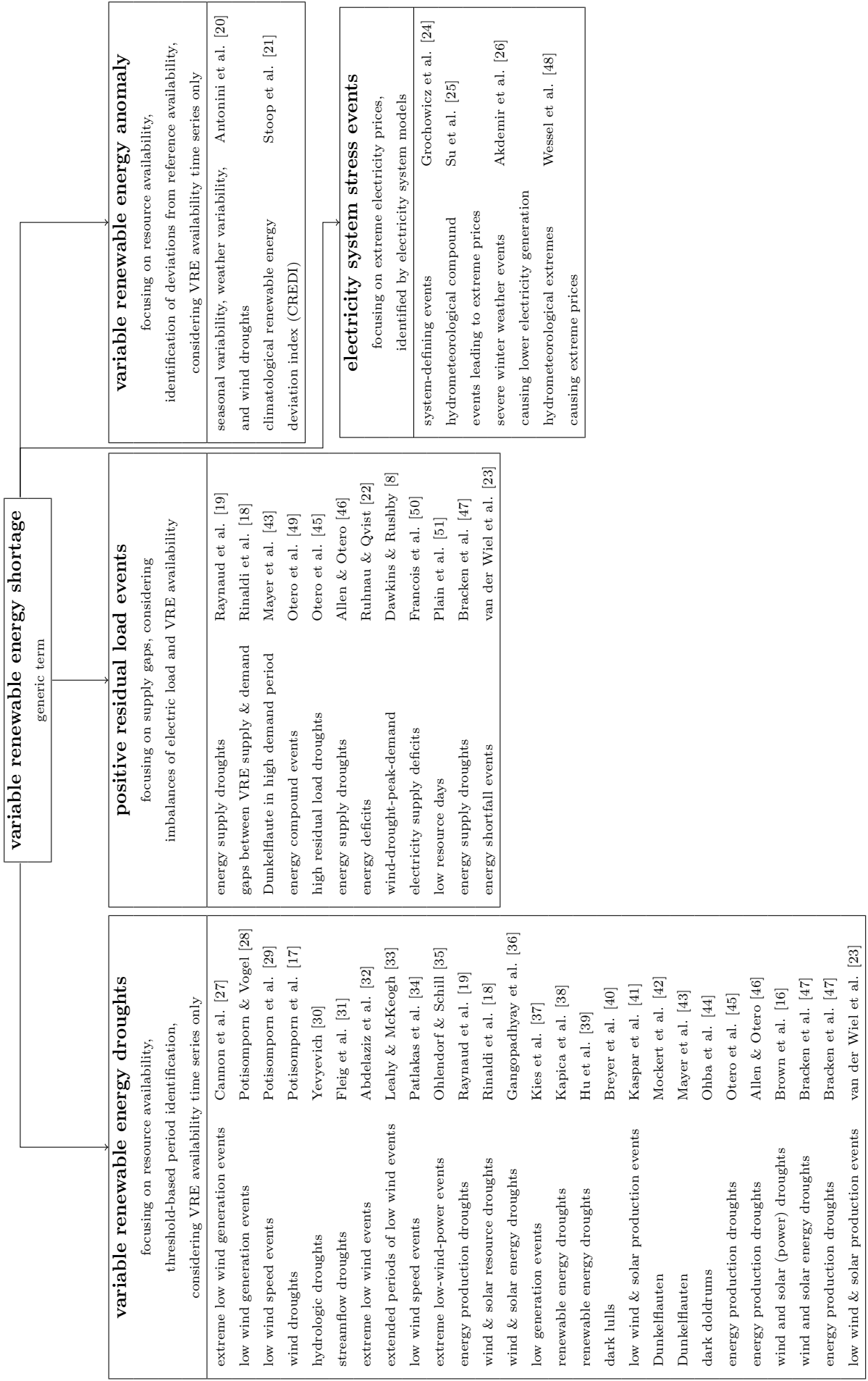
\begin{sidewaysfigure}
\scriptsize
\centering
\begin{tikzpicture}[] 
  \node (vreshortage) [rectangle, draw=none, align=center, text width=6.5cm] {
    };
  \node (vredrought) [below left=of vreshortage, xshift=-0.6cm, yshift=-0.5cm, rectangle, draw=none, align=center, text width=7cm] {
    };
  \node (prlevent) [below=of vreshortage, xshift=0cm, yshift=-0.5cm, rectangle, draw=none, align=center, text width=7cm] {
    };
  \node (anomaly) [below right=of vreshortage, xshift=0.33cm, yshift=-0.5cm, rectangle, draw=none, align=center, text width=7.2cm] {
    };
  \node (system_stress) [below=of anomaly, xshift=0cm, yshift=-3cm, rectangle, draw=none, align=center, text width=7.2cm] {
    };
  \node (table0) [below=-0.5cm of vreshortage, rectangle] {
    \begin{tabular}{|ll|}
      \hline
      \multicolumn{2}{|c|}{\normalsize{\textbf{variable renewable energy shortage}}}\\
      \multicolumn{2}{|c|}{\scriptsize{generic term}}\\
      \hline
    \end{tabular}
  };
  
  \draw[->] (vreshortage.west) -| (vredrought);
  \draw[->] ($(vreshortage.south) + (0,-0.48)$) -- (prlevent.north);
  \draw[->] (vreshortage.east) -| (anomaly);
  \draw[->] (vreshortage.east) -| ++(0.9cm, -4cm) |- ++(1.0cm, -1.5cm) -| (system_stress.north);

  \node (table1) [below=-0.3cm of vredrought, align=left] {
    \ra{1.05}
    \begin{tabular}{|ll|}
    \hline
    \multicolumn{2}{|c|}{\normalsize{\textbf{variable renewable energy droughts}}}\\
    \multicolumn{2}{|c|}{\scriptsize{focusing on resource availability,}}\\
    \multicolumn{2}{|c|}{\scriptsize{threshold-based period identification,}}\\
    \multicolumn{2}{|c|}{\scriptsize{considering \acs{VRE} availability time series only}}\\
    \hline
    extreme low wind generation events & Cannon et al. \cite{cannon_using_2015}\\
    low wind generation events & Potisomporn \& Vogel \cite{potisomporn_spatial_2022} \\
    low wind speed events & Potisomporn et al. \cite{potisomporn_evaluating_2023} \\
    wind droughts & Potisomporn et al. \cite{potisomporn_extreme_2024} \\
    hydrologic droughts & Yevyevich \cite{yevyevich_objective_1967} \\
    streamflow droughts & Fleig et al. \cite{fleig_global_2006} \\
    extreme low wind events & Abdelaziz et al. \cite{abdelaziz_assessing_2024} \\
    extended periods of low wind events & Leahy \& McKeogh \cite{leahy_persistence_2013} \\
    low wind speed events & Patlakas et al.~\cite{patlakas_low_2017} \\
    extreme low-wind-power events & Ohlendorf \& Schill \cite{ohlendorf_frequency_2020}\\
    energy production droughts & Raynaud et al. \cite{raynaud_energy_2018} \\
    wind \& solar resource droughts & Rinaldi et al. \cite{rinaldi_wind_2021} \\
    wind \& solar energy droughts & Gangopadhyay et al. \cite{gangopadhyay_role_2022} \\
    low generation events & Kies et al. \cite{kies_critical_2021}\\
    renewable energy droughts & Kapica et al. \cite{kapica_potential_2024} \\
    renewable energy droughts & Hu et al. \cite{hu_implications_2023} \\
    dark lulls & Breyer et al. \cite{breyer_reflecting_2022} \\
    low wind \& solar production events & Kaspar et al. \cite{kaspar_climatological_2019} \\
    Dunkelflauten & Mockert et al. \cite{mockert_meteorological_2023} \\
    Dunkelflauten & Mayer et al. \cite{mayer_probabilistic_2023} \\
    dark doldrums & Ohba et al. \cite{ohba_climatology_2022} \\
    energy production droughts & Otero et al. \cite{otero_copula-based_2022} \\
    energy production droughts & Allen \& Otero \cite{allen_standardised_2023} \\
    wind and solar (power) droughts & Brown et al. \cite{brown_meteorology_2021} \\
    wind and solar energy droughts & Bracken et al. \cite{bracken_standardized_2024}\\
    energy production droughts & Bracken et al. \cite{bracken_standardized_2024} \\
    low wind \& solar production events & van der Wiel et al. \cite{van_der_wiel_influence_2019} \\
    \hline
    \end{tabular}
  };
  \node (table3) [below=-0.3cm of anomaly, align=left] {
    \ra{1.05}
    \begin{tabular}{|ll|}
    \hline
    \multicolumn{2}{|c|}{\normalsize{\textbf{variable renewable energy anomaly}}}\\
    \multicolumn{2}{|c|}{\scriptsize{focusing on resource availability,}}\\
    \multicolumn{2}{|c|}{\scriptsize{identification of deviations from reference availability,}}\\
    \multicolumn{2}{|c|}{\scriptsize{considering \acs{VRE} availability time series only}}\\
    \hline
    seasonal variability, weather variability, & Antonini et al. \cite{antonini_identification_2024} \\
    and wind droughts & \\
    climatological renewable energy & Stoop et al. \cite{stoop_climatological_2024}\\
    deviation index (CREDI) & \\
    \hline
    \end{tabular}
  };
  \node (table4) [below=-0.3cm of system_stress, align=left] {
    \ra{1.05}
    \begin{tabular}{|ll|}
    \hline
    \multicolumn{2}{|c|}{\normalsize{\textbf{electricity system stress events}}}\\
    \multicolumn{2}{|c|}{\scriptsize{focusing on extreme electricity prices,}}\\
    \multicolumn{2}{|c|}{\scriptsize{identified by electricity system models}}\\
    \hline
    system-defining events & Grochowicz et al. \cite{grochowicz_using_2024} \\
    hydrometeorological compound & Su et al. \cite{su_compound_2020} \\
    events leading to extreme prices & \\
    severe winter weather events & Akdemir et al. \cite{akdemir_assessing_2022}\\
    causing lower electricity generation & \\
    hydrometeorological extremes & Wessel et al. \cite{wessel_technology_2022} \\
    causing extreme prices & \\
    \hline
    \end{tabular}
  };
  \node (table2) [below=-0.3cm of prlevent, align=left] {
    \ra{1.05}
    \begin{tabular}{|ll|}
    \hline
    \multicolumn{2}{|c|}{\normalsize{\textbf{positive residual load events}}}\\
    \multicolumn{2}{|c|}{\scriptsize{focusing on supply gaps, considering}}\\
    \multicolumn{2}{|c|}{\scriptsize{imbalances of electric load and \acs{VRE} availability}}\\
    \multicolumn{2}{|c|}{}\\
    \hline
    energy supply droughts & Raynaud et al. \cite{raynaud_energy_2018}\\
    gaps between \acs{VRE} supply \& demand & Rinaldi et al. \cite{rinaldi_wind_2021} \\
    Dunkelflaute in high demand period & Mayer et al. \cite{mayer_probabilistic_2023} \\
    energy compound events & Otero et al. \cite{otero_characterizing_2022} \\
    high residual load droughts & Otero et al. \cite{otero_copula-based_2022} \\
    energy supply droughts & Allen \& Otero \cite{allen_standardised_2023} \\
    energy deficits & Ruhnau \& Qvist \cite{ruhnau_storage_2022} \\
    wind-drought-peak-demand & Dawkins \& Rushby \cite{dawkins_characterising_2020} \\
    electricity supply deficits & Francois et al. \cite{francois_statistical_2022} \\
    low resource days & Plain et al. \cite{plain_accounting_2019} \\
    energy supply droughts & Bracken et al. \cite{bracken_standardized_2024}\\
    energy shortfall events & van der Wiel et al. \cite{van_der_wiel_influence_2019} \\
    \hline
    \end{tabular}
  };
\end{tikzpicture}
\caption{Proposal for a distinctive terminology and overview of previously used terms in two major and one emerging strand of \ac{VRE} shortage analysis.}
\label{fig:terminology}
\end{sidewaysfigure}

\ac{VRE} shortage can be analyzed using different types of input data. One approach that focuses on energy supply is to detect periods with low renewable availability. The latter can be derived from measurements, modeled weather variables, or renewable availability time series, recognizing technological assumptions and spatial distribution of the installed capacity. Shortage can be detected by identifying concrete periods with availability below a given threshold. For this, many different terms have been used in the literature (Figure \ref{fig:terminology}), e.g., ``wind drought'' \cite{potisomporn_extreme_2024}, ``wind and solar resource drought'' \cite{rinaldi_wind_2021}, or ``energy production droughts'' \cite{raynaud_energy_2018} in the literature. An alternative is to compute deviations of renewable availability from a reference profile over a user-specific time horizon. The latter has been termed  ``seasonal variability'',  ``weather variability'', or  ``wind droughts'' \cite{antonini_identification_2024}, as well as ``cumulative anomaly of a renewable resource'' \cite{stoop_climatological_2024}.

Another approach is to consider not only \ac{VRE} generation potentials, but also how these relate to electric load profiles. This can be done by analyzing residual load, sometimes also referred to as net load, which represents the net balance of electricity demand and \ac{VRE} supply. Again, there is no consensus in the literature on how to term such shortage events (Figure \ref{fig:terminology}), which, for example, have been referred to as ``energy supply droughts'' \cite{raynaud_energy_2018}, ``energy deficits'' \cite{ruhnau_storage_2022}, or ``energy shortfall events'' \cite{van_der_wiel_influence_2019}. 

Alternatively, power system model outcomes that use \ac{VRE} availability as input can be used to identify extreme \ac{VRE} conditions causing power system stress. Focusing on high electricity prices, corresponding periods have been termed as ``system-defining'' \cite{grochowicz_using_2024}, ``hydrometeorological compound events'' \cite{su_compound_2020}, or ``severe winter weather events'' \cite{akdemir_assessing_2022}.

For clarity, we propose the following terminology: ``variable renewable energy droughts'' for periods with low resource availability of a single or multiple \ac{VRE} technologies, ``variable renewable energy anomaly'' for cumulative deviations of \ac{VRE} resource availability from a given mean reference, ``\ac{PRL} events'' for periods where variable renewable supply falls short of electric load, and ``electricity system stress events'' for periods with high electricity prices signaling extreme renewable availability. The generic term ``variable renewable energy shortage''  captures these concepts (Figure~\ref{fig:terminology}). Note that \ac{VRE} drought and \ac{PRL} events refer to concrete periods and do not necessarily occur at the same time. While both concepts characterize variable renewable energy shortage, \ac{PRL} events are additionally affected by the timing of coinciding high-demand periods. In contrast, \ac{VRE} anomalies are measured over user-specific time horizons. Electricity system stress events may be driven by further technical limitations such as grid constraints, impacting their occurrence patterns.

\ac{VRE} shortage analyses can be carried out with various methods, ranging from stylized, assumption-based projections of future energy mixes to more sophisticated, optimizing energy system models. Further, the scope of variable renewable energy shortage analyses is broad, ranging from the assessment of individual renewable technologies in specific locations to optimized portfolios of generation technologies and demand flexibility across large balancing areas, encompassing multiple countries. Consequently, the definitions and identification methods for \ac{VRE} shortage events should facilitate meaningful comparisons across various renewable generation technologies and regions.

Different bodies of literature contribute to the analysis of \ac{VRE} shortage events, primarily related to meteorology or energy research. One literature strand focuses on wind energy. Notably, wind droughts in the United Kingdom are well-studied. Cannon et al. \cite{cannon_using_2015} explore extreme onshore wind power events in the UK based on reanalysis data. Thresholds correspond to percentiles of the cumulative frequency distribution of investigated data. Potisomporn \& Vogel \cite{potisomporn_spatial_2022} study the spatial and temporal correlation of low-wind-power events for offshore wind locations in the United Kingdom given a historical and projected distribution of offshore wind farms. Potisomporn et al. \cite{potisomporn_evaluating_2023} contrast low-wind speed events derived from ERA5 reanalysis data to measured data from 205 onshore and offshore wind locations in the United Kingdom. The same authors in \cite{potisomporn_extreme_2024} compare different identification methods for low-wind events in the United Kingdom based on reanalysis data, deploying concepts for analyzing hydrological droughts \cite{yevyevich_objective_1967,fleig_global_2006}. Abdelaziz et al. \cite{abdelaziz_assessing_2024} assess offshore wind speed droughts in the United Kingdom. They identify consecutive days with daily average wind speed below a given threshold. There are related studies analyzing other regions. Leahy \& McKeogh \cite{leahy_persistence_2013} and Patlakas et al.~\cite{patlakas_low_2017} study the persistence of low-wind speeds for different locations in Ireland and the North Sea based on historical records. They search for events with wind speeds constantly below a threshold and for annual minimal values of moving average wind speeds over pre-defined intervals. Thresholds are based on a range of exogenously set wind speeds. Ohlendorf \& Schill \cite{ohlendorf_frequency_2020} investigate the duration and frequency of low onshore wind power events in Germany based on reanalysis data and exogenously set thresholds.

While the first \ac{VRE} drought analyses focused on wind power, research interest in droughts of variable renewable energy portfolios that also include solar \ac{PV} is growing. Raynaud et al. \cite{raynaud_energy_2018} investigate droughts for wind, solar, and hydro, both separately and combined as a portfolio, for 12 European regions. They employ relative regional thresholds as fractions of the mean daily availability factor. Rinaldi et al. \cite{rinaldi_wind_2021} draw on Raynaud et al. \cite{raynaud_energy_2018} for identifying wind, solar, and wind-solar droughts in the Western U.S. for different thresholds relative to mean daily generation. Gangopadhyay et al. \cite{gangopadhyay_role_2022} apply a similar approach to India, drawing on 5000 years of synthetic data. Kies et al. \cite{kies_critical_2021} compare seven \ac{VRE} availability datasets for Europe. Using a composite \ac{VRE} portfolio time series that combines wind and \ac{PV} availability, they search for periods constantly below a range of exogenous thresholds. Kapica et al. \cite{kapica_potential_2024} compare energy droughts for solar \ac{PV} and wind in Europe, either separately or as a combined portfolio, for historical and future weather conditions retrieved from eight regional climate models. Their identification method again draws on \cite{raynaud_energy_2018} with thresholds relative to seasonal and regional averages. Hu et al. \cite{hu_implications_2023} explore the risk of simultaneously occurring \ac{VRE} droughts for European countries for both single technologies and a \ac{VRE} portfolio, and compare it to droughts arising across all of Europe assuming perfect electricity exchange. Breyer et al. \cite{breyer_reflecting_2022} model decarbonization pathways for Europe until 2040 or 2050, using the weather year 2005. They screen their model outcomes for drought periods constantly below a range of thresholds relative to the maximum availability of wind and \ac{PV} separately and combined for Germany and Europe. Kaspar et al. \cite{kaspar_climatological_2019} analyze wind and \ac{PV} separately and combined for Germany. For given averaging intervals, they scan for moving averages below exogenously set thresholds. Mockert et al. \cite{mockert_meteorological_2023} also study \ac{VRE} droughts in Germany in the context of prevailing weather regimes and lean on Kaspar et al. \cite{kaspar_climatological_2019} for drought identification. Mayer et al. \cite{mayer_probabilistic_2023} analyze concurrent \ac{PV} and wind droughts in Hungary by identifying consecutive time steps with availability from both technologies below exogenous thresholds. Ohba et al. \cite{ohba_climatology_2022} link wind and solar droughts, both separately and combined as a portfolio, to four typical weather patterns arising in eastern Japan using an exogenously set threshold and neural network learning.

In another strand of the literature, \ac{VRE} shortage events are analyzed in the context of residual load. In addition to their \ac{VRE} drought analysis, Raynaud et al. \cite{raynaud_energy_2018} investigate the imbalance of demand and \ac{VRE} supply in Europe. Van der Wiel et al. \cite{van_der_wiel_influence_2019} test whether low wind and solar generation as well as high residual events are linked to large-scale weather regimes arising over Europe. Otero et al. \cite{otero_characterizing_2022} investigate periods with simultaneously low wind and solar generation and high demand across 27 European countries based on absolute thresholds. Otera et al. \cite{otero_copula-based_2022} use copula analysis to analyze \ac{VRE} droughts based on a threshold relative to a ten percent fraction of its mean availability, and residual load events above the 90\% percentile of its distribution in Europe based on ERA5 reanalysis data. Allen \& Otero \cite{allen_standardised_2023} introduce standardized \ac{VRE} droughts indices derived from meteorological indicators that incorporate both \ac{VRE} generation and electricity demand. They showcase the proposed method for droughts in Europe. Ruhnau \& Qvist \cite{ruhnau_storage_2022} retrieve residual load time series from a model analysis of a least-cost decarbonized Germany power sector scenario and search for persistent positive residual load events. Dawkins \& Rushby \cite{dawkins_characterising_2020} characterize wind droughts and residual load events that qualify as droughts in the United Kingdom based on percentile thresholds. Francois et al. \cite{francois_statistical_2022} explore positive residual load events (coined electricity deficits) for run-of-river hydropower and solar \ac{PV} net of demand for Northern Italy, focusing on return periods lasting from single hours up to several years. Based on the assumption of sufficient storage availability, Plain et al. \cite{plain_accounting_2019} identify low-solar periods using satellite data to size microgrids power systems in Africa based on the mismatch between solar production and demand. Bracken et al. \cite{bracken_standardized_2024} characterize \ac{VRE} droughts across the U.S. concerning frequency, duration, magnitude, and seasonality given the existing energy infrastructure in 2020. They investigate single \ac{VRE} technologies, \ac{VRE} portfolios, and positive residual load events.

Recently, another literature strand has been emerging in which cumulative \ac{VRE} anomalies from some reference over a given time horizon are identified. Antonini et al. \cite{antonini_identification_2024} identify wind energy droughts on a global scale as deviation of hourly wind power density from its multidecadal, climatological mean hourly profile. They identify world well-suited wind regions characterized by high wind power density, low inter- and intra-annual variability, and less pronounced droughts. Stoop et al.~\cite{stoop_climatological_2024} follow a similar approach and introduce a climatological renewable energy deviation index (CREDI), which quantifies cumulative anomalies of renewable resources over different timescales, ranging from hours to decades. While this method is not limited to analyses of renewable energy shortage events, the authors present an exemplary application for 8-day Dunkelflaute events in the paper. The approaches in this emerging literature strand avoid some of the methodological challenges of \ac{VRE} availability and \ac{PRL} event analysis that will be discussed below. Yet, the choice of the time horizon remains user-specific. Further, it is not clear how the chosen reference profiles that are used to compute renewable anomalies relate to energy system needs.

Alternatively, the effect of extreme \ac{VRE} conditions causing power system stress events can be detected based on electricity or energy system model outcomes. Grochowicz et al.~\cite{grochowicz_using_2024} detect system stress events in Europe based on electricity shadow prices derived from a power sector model. They identify system-defining events that may last up to two weeks and account for a considerable share of total system costs. Such events occur between November and February and last between two and 13 days. Importantly, their findings are only applicable to the particular model setting and rely on a range of model-inherent parameter assumptions. Similar analyses have been carried out for different regions in the U.S., focusing on electricity prices and emissions \cite{su_compound_2020}, electricity prices and wind generation during winter storms \cite{akdemir_assessing_2022} as well as electricity prices and system stability \cite{wessel_technology_2022}.

The review shows that different bodies of literature, e.g., hydrology, wind and solar energy, or energy system modeling, use a wide range of methodological approaches for defining and quantifying \ac{VRE} shortage. While alternative approaches are emerging, the vast majority of the reviewed studies use threshold-based methods for analyzing variable renewable energy drought and positive residual load events (in the latter case, using a zero threshold). Additionally, existing studies differ in terms of the technological, geographical, and temporal scope of the investigated system, based on partly inconsistent terminology. This could be a result of the disconnect of involved research fields. Communities at the intersections of these previously disconnected fields have established, bridging the gap between meteorology and energy research \cite{bloomfield_importance_2021,craig_overcoming_2022}. Yet, the literature lacks a structured overview and explanation of different methodological approaches for \ac{VRE} shortage analyses using harmonized terminology. We aim to contribute to the literature by closing this gap.

While we do not aim to provide a complete review of the existing literature, we provide a structured overview of different approaches for identifying variable renewable energy droughts across multiple regions and technologies from the perspective of the energy system. Based on the literature review, we focus our analysis on threshold-based methods, including variable renewable energy droughts and positive residual load events. We discuss possible definitions of \ac{VRE} droughts and identification methods, focusing on renewable availability time series. As variable renewable energy potentials vary across technologies, space, and time, we elaborate on how to enable comparability across these dimensions, which has not yet been considered in the literature. We further discuss how \ac{VRE} drought analysis can be extended to also include electric load, i.e.,~how to evaluate extreme \ac{PRL}. This allows for identifying compound low-\ac{VRE} availability and high-demand events indicating the need for dispatchable capacity. Drawing on the proposed method for \ac{VRE} availability time series, we discuss how to analyze \ac{PRL} events and discuss essential points of criticism.

\section{Concepts for defining variable renewable energy droughts}\label{sec:vre_droughts_supply}

\subsection{Drought definitions}

When exploring \ac{VRE} droughts, i.e., shortage on the energy supply side, one approach is to detect periods with low wind speeds or solar irradiation. According data can originate from measurements (in-situ or by satellites) or meteorological models. Alternatively, drought analyses can use \ac{VRE} availability time series, also referred to as capacity factors, either derived from measurements or converted from modeled weather variables. These indicate how much electricity (e.g., in megawatt hours) a renewable source can generate at a given time step $t$, normalized by its installed generation capacity (e.g., in megawatt). In other words, an availability factor is a measure of a \ac{VRE} generator's capture efficiency of the meteorological resource \cite{nuno_simulation_2018}. While our analysis focuses on discrete time series of \ac{VRE} availability factors $avail_t \in [0, 1]$, the discussed methods generally apply to all of the above-mentioned data types. For clarity, we start with a single technology and one region, and later turn to renewable technology portfolios and larger regions.

\begin{figure}[htbp]
\centering
\subfloat[Drought windows with fixed duration below a drought threshold $thres = 0.1$.\label{fig:window_0.1}]
    {{\includegraphics[width=.97\textwidth]{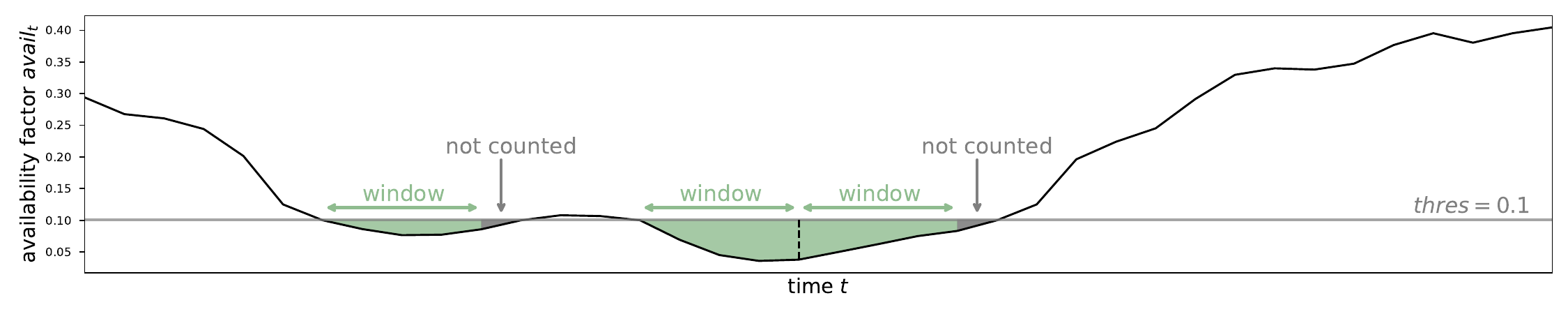}}}
\qquad
\subfloat[Drought events with variable duration below two drought thresholds $thres_1 = 0.1$ and $thres_1 = 0.2$. Using higher thresholds generally leads to longer, but fewer, events.\label{fig:event}]
    {{\includegraphics[width=.97\textwidth]{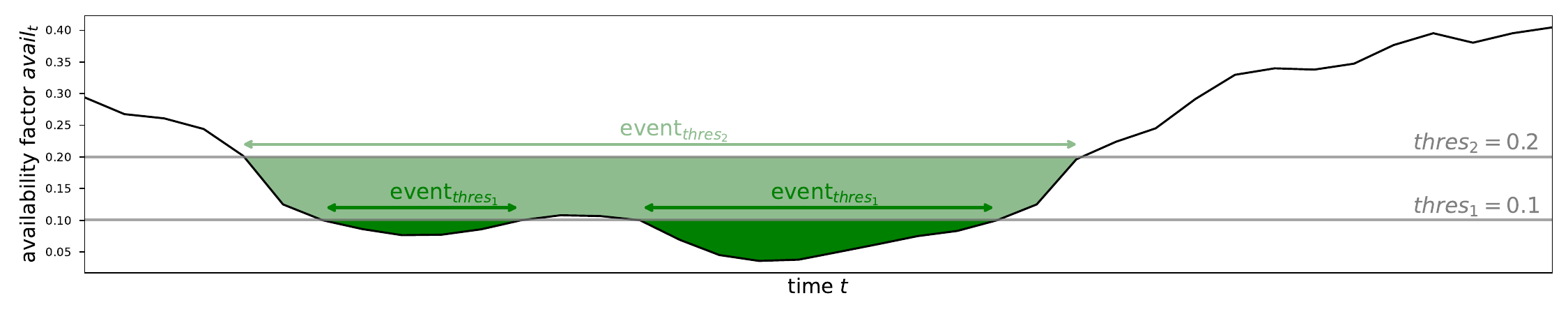}}}\\
\caption{Comparison of \ac{VRE} drought definitions. The window perspective underestimates drought duration and overestimates the number of droughts.}%
\label{fig:window_event}%
\end{figure}

Given a \ac{VRE} availability time series, variable renewable energy droughts can generally be defined as periods below a certain drought qualification threshold $thres \in [0, 1]$ (gray lines in Figure \ref{fig:window_event}). There are different definitions for \ac{VRE} drought periods. One option is searching for 'drought windows' with a predefined duration \cite{raynaud_energy_2018,kapica_potential_2024,rinaldi_wind_2021}\footnote{These studies search for periods of consecutive daily averages below some threshold with a cut-off every 24 hours.}, e.g., a given number of time steps below a given threshold (Figure \ref{fig:window_0.1}). Implementing a respective search algorithm for a specified window duration is straightforward. Yet, droughts identified via predefined windows may be shorter than actual droughts (Figure~\ref{fig:window_0.1}). Further, consecutive windows that form one longer combined drought may be counted individually (Figure~\ref{fig:window_0.1}, right windows). Thus, this definition systematically underestimates drought duration and overestimates the number of droughts.

An alternative approach is searching for complete 'drought events' below a threshold, which last as long as a drought qualification criterion is met. That is, drought duration is not predefined by setting a window length, but, for example, the result of a search on how many consecutive time steps are below a given threshold \cite{yevyevich_objective_1967,fleig_global_2006,potisomporn_extreme_2024}. This ensures that identified events are completely captured, and consecutive droughts are properly accounted for (Figure~\ref{fig:event}). We thus consider the drought event perspective to be a superior concept for identifying \ac{VRE} droughts, and focus on it henceforward.

Higher thresholds generally identify less severe, but longer-lasting drought events. This tends to reduce the number of obtained droughts, as events that are considered to be separate for lower thresholds may integrate into combined ones (Figure~\ref{fig:event}).

\subsection{Drought identification methods}\label{ssec:vre_cbt_mbt_spc}
 
\begin{figure}[htbp]
\centering
\noindent\includegraphics[width=.97\linewidth, keepaspectratio]{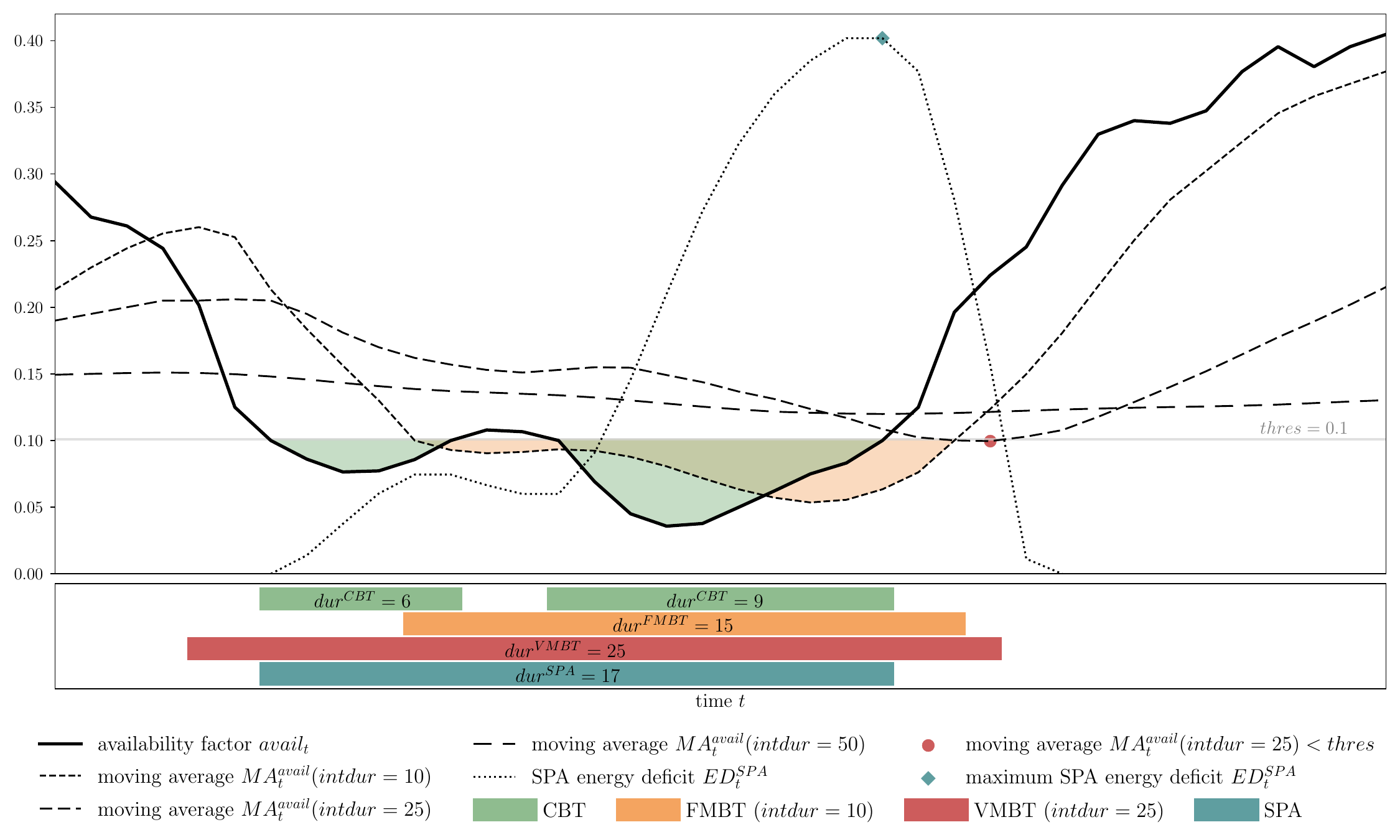}
\caption{The upper panel shows exemplary availability time series (bold line) and identification methods \acf{CBT}, \ac{FMBT}, \ac{VMBT}, and \ac{SPA}. The gray line represents the drought threshold $thres = 0.1$. Green and orange areas highlight the energy deficit obtained by \ac{CBT} and \ac{FMBT}, respectively. The lower panel's horizontal bars illustrate the varying event duration $dur$ identified by each method.}
\label{fig:cbt_mbt_spc}
\end{figure}

Numerous methods exist for identifying \ac{VRE} droughts. We focus on a selection\footnote{We limit our selection to methods that appear useful for investigating \ac{VRE} droughts. Other methods in different bodies of literature can be found, e.g, the Inter-event time method that was developed for hydropower \cite{zelenhasic_method_1987} and was recently evaluated in \cite{potisomporn_extreme_2024}.} in the following (Figure \ref{fig:cbt_mbt_spc}). A \ac{CBT} event relates to a period of consecutive time steps during which availability factors are constantly below a threshold $thres$ \cite{guerrero-salazar_analysis_1975,leahy_persistence_2013,cannon_using_2015,patlakas_low_2017,ohlendorf_frequency_2020,potisomporn_spatial_2022,potisomporn_extreme_2024,mayer_probabilistic_2023}. In the hydrologic literature, this method has been first introduced as 'runs analysis' \cite{yevyevich_objective_1967}. The boolean function $VRED_t^{CBT}$ determines whether time step $t$ qualifies as a variable renewable energy drought (1) or not~(0):

\begin{equation}
    VRED_t^{CBT} = 
    \begin{cases}
        1 & \text{if $avail_t < thres \ \ \forall \ \ t$}\\
        0 & \text{if $avail_t \geq thres \ \ \forall \ \ t$}
    \end{cases}
\end{equation}

The duration of a \ac{CBT} event $dur^{CBT}$ is the number of consecutive time steps with $VRED_t^{CBT} = 1$. Suppose $t_k$ and $t_l$ are the first and the last time steps of an event. Then the event's cumulative energy, also referred to as energy deficit or drought severity \cite{yevyevich_objective_1967,potisomporn_extreme_2024}, is defined as follows: 

\begin{equation}
    ED_t^{CBT} = \sum_{t_k}^{t_l} thres - avail_t \ \ \forall \ \ t_k \leq t \leq t_l
\end{equation}

The \ac{CBT} notion identifies \ac{VRE} droughts regardless of availability factors in adjacent periods, even if their availability is briefly above the drought threshold (Figure \ref{fig:cbt_mbt_spc}).

Addressing this limitation, \ac{MBT} events define consecutive time steps during which a moving average of availability factors $MA_t^{avail}(intdur)$ is below a threshold \cite{raynaud_energy_2018,ohlendorf_frequency_2020,kaspar_climatological_2019,mockert_meteorological_2023,potisomporn_extreme_2024}. This approach smooths out short-term fluctuations over the lagging averaging interval $[t-intdur, t]$\footnote{We propose using a lagging averaging interval, which is is common in time series analyses. Using a leading interval would also be possible, but appears non-intuitive. Likewise, a centered interval is also not recommended as dealing with border time steps for an uneven number of time steps is ambiguous.} with a predefined duration $intdur$. Equivalent to $VRED_t^{CBT}$, the function $VRED_t^{MBT}$ evaluates each time step $t$ as follows:

\begin{equation}
    VRED_t^{MBT} = 
    \begin{cases}
        1 & \text{if $MA_t^{avail}(intdur) < thres \ \ \forall \ \ t$}\\
        0 & \text{if $MA_t^{avail}(intdur) \geq thres \ \ \forall \ \ t$}
    \end{cases}
\end{equation}

The \ac{MBT} method can be applied in two different ways. First, the \acf{FMBT} searches for \ac{MBT} events based on a fixed averaging interval duration $intdur^{FMBT}$ \cite{kaspar_climatological_2019,mockert_meteorological_2023,potisomporn_extreme_2024}. The duration of an \ac{FMBT} event $dur^{FMBT}$ corresponds to the number of consecutive time steps with $VRED_t^{FMBT}=1$ (e.g., $dur^{FMBT}=15$ for $intdur=10$ in Figure~\ref{fig:cbt_mbt_spc}). The energy deficit $ED_t^{FMBT}$ is computed analogously to $ED_t^{CBT}$ (i.e., using $MA_t^{avail}(intdur)$ instead of $avail_t$) and decreases due to the smoothing effect of the moving average. One option to avoid this is evaluating the energy deficit based on the original time series $avail_t$ instead of $MA_t^{avail}(intdur)$, i.e., reporting $ED_t^{CBT}$ instead of $ED_t^{FMBT}$ corresponding to the identified \ac{FMBT} event \cite{potisomporn_extreme_2024}.

Unlike \ac{CBT}, the \ac{FMBT} approach allows for instances above the drought threshold before, between, and after potentially multiple \ac{CBT} drought events that are filtered out by pooling them together (compare \ac{CBT} and \ac{FMBT} events in Figure~\ref{fig:cbt_mbt_spc}) \cite{potisomporn_extreme_2024}. In the hydrology literature, such \ac{CBT} events are referred to as 'mutually dependent' \cite{fleig_global_2006}. For prolonging averaging intervals, the moving average smoothing becomes more pronounced. Hence, the choice of $intdur^{FMBT}$ determines to which extent shorter \ac{CBT} events are filtered out.\footnote{The \ac{FMBT} method converges to \ac{CBT} as the averaging interval duration $intdur^{FMBT}$ approaches the granularity of the original time series.} By design, for a given threshold, the \ac{CBT} approach identifies more drought events than \ac{FMBT}, whereas \ac{FMBT} finds fewer but longer-lasting events. Thus, the \ac{CBT} method may overestimate the number of drought events while underestimating the duration and energy deficit, which is particularly relevant for characterizing the cumulative energy of extreme \ac{VRE} droughts. In contrast, the \ac{FMBT} method acknowledges natural fluctuations in \ac{VRE} availability by accounting for brief periods of higher availability within an overall low-availability period. \ac{FMBT} therefore provides a more realistic view of prolonged \ac{VRE} droughts and the corresponding power sector flexibility needs.

However, there is no clear relation between a power sector attribute and $intdur^{FMBT}$, whose parameterization remains arbitrary. \ac{VRE} availability time series could be evaluated for a range of different $intdur^{FMBT}$. Yet, drought events based on different $intdur^{FMBT}$ may partially overlap, creating ambiguity in their characteristics like duration, energy deficit, and frequency. In fact, results may substantially change with varying $intdur^{FMBT}$ (compare the moving averages $MA_t^{avail}(intdur = 10)$ and $MA_t^{avail}(intdur=50)$ in Figure \ref{fig:cbt_mbt_spc}, with the latter obtaining no drought at all). Consequently, \ac{VRE} droughts identified via the \ac{FMBT} approach are valid only for a specific parameterization of $intdur^{FMBT}$, limiting the applicability of this concept.

Addressing this limitation, we introduce the \acf{VMBT} method, which to the best of our knowledge is a novelty. It obtains unique \ac{VRE} droughts, ranging from very long-lasting events, e.g., months, to such that last only a few time steps. This is achieved by deploying the \ac{MBT} perspective for a large number of different averaging intervals, iteratively adjusting $intdur^{VMBT}$ in descending order. Importantly, to ensure that no long-duration events with $dur^{VMBT} > intdur^{VMBT}$ are missed, the first iteration should start with a very high $intdur^{VMBT}$, so that the moving average $MA_t^{avail}(intdur^{VMBT}) > thres$ across the entire time series (in the example shown in Figure~\ref{fig:cbt_mbt_spc}, compare the moving average $MA_t^{avail}(intdur = 50)$). In subsequent iterations, $intdur^{VMBT}$ is reduced and $MA_t^{avail}$ updated, until the first drought event is identified at $MA_t^{avail}(intdur^{VMBT}) < thres$ (compare the moving average $MA_t^{avail}(intdur = 25)$ in Figure~\ref{fig:cbt_mbt_spc}). Its duration equals the averaging interval duration $dur^{VMBT} = intdur^{VMBT}$. The respective period is then excluded in later iterations, where $intdur^{VMBT}$ decreases further and additional (shorter) drought events are identified, and so on. For a given drought threshold $thres$, this iterative process effectively captures all unique drought events, without any overlap or double counting, each accurately recorded with its maximum possible duration. Compared to \ac{FMBT}, this renders the \ac{VMBT} approach a more accurate and informative method for drought identification.

Note that the \ac{VMBT} energy deficit $ED_t^{VMBT}$ is insignificant due to the close proximity of the moving average to the threshold. Further, this method involves a trade-off between accuracy and computational efficiency. Iteratively reducing the interval duration of the moving average ($intdur^{VMBT}$), with an incremental change that equals the resolution of the underlying time series, ensures captures all unique drought events, but requires numerous iterations and may lead to long processing time.

Finally, the \acf{SPA} identification method searches for the maximum cumulative energy deficit of a drought event, while accommodating intermediate periods where renewable availability may surpass the threshold. Originating in hydrology \cite{vogel_generalized_1987,fleig_global_2006}, \ac{SPA} has been extended to other bodies of literature, such as wind energy analysis \cite{potisomporn_extreme_2024}. A similar concept has also been applied to characterizing renewable surplus events \cite{schill_residual_2014}. The cumulative energy deficit is defined as follows:

\begin{equation}
    ED_t^{SPA} = 
    \begin{cases}
        ED^{SPA}_{t-1} + thres -  avail_t & \text{if $ED^{SPA}_{t-1} + thres -  avail_t > 0 \ \ \forall \ \ t$}\\
        0 & \text{if $ED^{SPA}_{t-1} + thres -  avail_t \leq 0 \ \ \forall \ \ t$}
    \end{cases}
\end{equation}

The \ac{SPA} measures event duration $dur^{SPA}$ from the first instance when the cumulative energy deficit of a single \ac{SPA} event becomes positive to its global\footnote{Here, ``global'' refers to the maximum of a single \ac{SPA} event, not to the maximum of all potential SPA events of the underlying \ac{VRE} availability time series.} maximum. This maximum occurs when the \ac{VRE} availability curve intersects with the threshold for the last time during an \ac{SPA} event, i.e.,~when $avail_t \geq thres$ (Figure~\ref{fig:cbt_mbt_spc}). This algorithm requires no iterations or assumptions on averaging intervals, and only a single straightforward computation of cumulative energy deficits for the underlying \ac{VRE} availability time series. It allows for long-duration events that include brief periods with above-threshold availability.

Unlike \ac{VMBT}, \ac{SPA} events start when \ac{VRE} availability falls below the threshold and thus depends on threshold parameterization. In contrast, the \ac{VMBT} method allows for high or low availability periods before \ac{SPA} events, making it less reliant on specific threshold parameterization. For a given threshold, both \ac{VMBT} and \ac{SPA} obtain a unique and unambiguous set of \ac{VRE} drought events, and thus appear to be superior to the \ac{CBT} of \ac{FMBT} approaches. \ac{SPA} is computationally less demanding than \ac{VMBT}, as it requires no iterations over averaging intervals. Yet, \ac{SPA} may not recognize minor events that follow a longer-lasting drought, which has been criticized in the literature \cite{potisomporn_extreme_2024}.

As indicated by Figure~\ref{fig:cbt_mbt_spc}, for a given threshold, the \ac{VMBT} approach generally leads to the longest drought duration of all four identification strategies explained above. The \ac{SPA} method identifies events with extended duration, albeit shorter than \ac{VMBT}, followed by \ac{FMBT} events with moderate duration. On the other end of the spectrum, the \ac{CBT} method leads to the shortest drought duration.

While often reported in other bodies of literature, energy deficits of \ac{VRE} droughts identified purely based on \ac{VRE} availability are hard to interpret as they do not allow for drawing direct energy system conclusions.

Note that an effective threshold parameterization is methodologically challenging and subject to user discretion (see Section~\ref{ssec:vre_scaling} for further discussion). 

\subsection{Comparability requirements and the choice of threshold}\label{ssec:vre_scaling}

Research interest is increasingly moving beyond single technologies at particular regions (e.g., \cite{cannon_using_2015,potisomporn_spatial_2022,potisomporn_evaluating_2023,abdelaziz_assessing_2024,ohlendorf_frequency_2020}) towards a characterization of \ac{VRE} drought patterns of portfolios of different renewable generation technologies for multiple regions (e.g., \cite{raynaud_energy_2018,rinaldi_wind_2021,kies_critical_2021,kapica_potential_2024}). For this, enabling comparability across \ac{VRE} technologies and regions when analyzing \ac{VRE} drought events is imperative.

\begin{figure}[htbp]
\centering
\subfloat[Availability duration curves of single \ac{VRE} technologies.\label{fig:duration_curves_single}]
    {{\includegraphics[width=.47\textwidth]{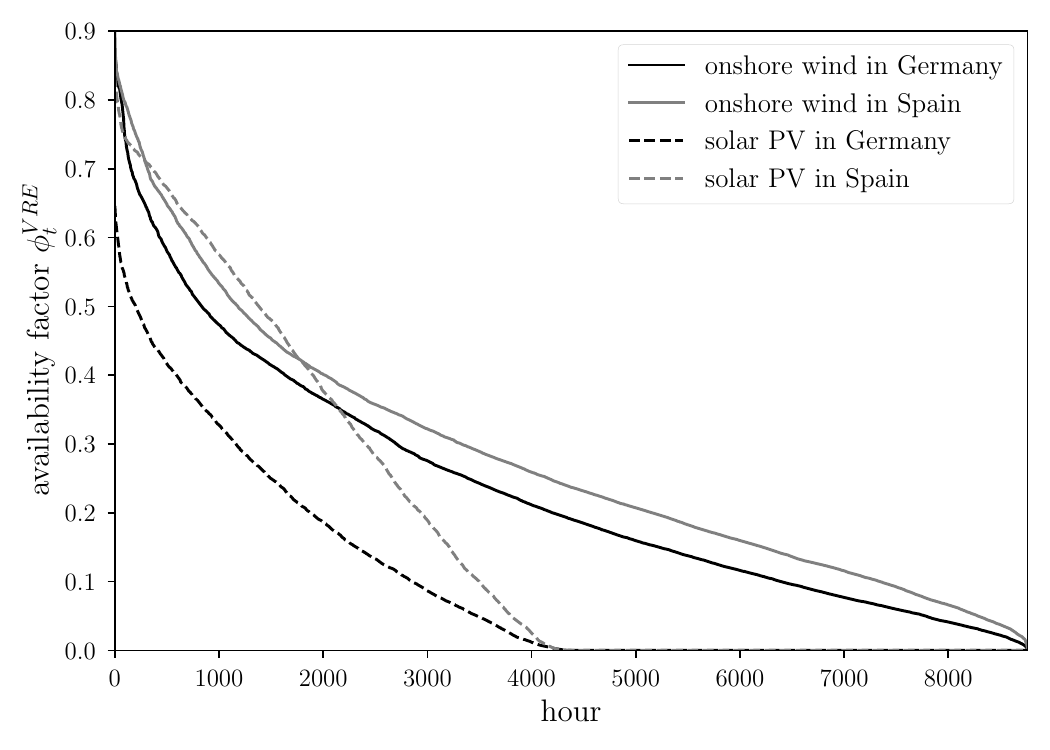}}}
\qquad
\subfloat[Availability duration curves of \ac{VRE} portfolios with different exemplary capacity mixes.\label{fig:duration_curves_portfolio}]
    {{\includegraphics[width=.47\textwidth]{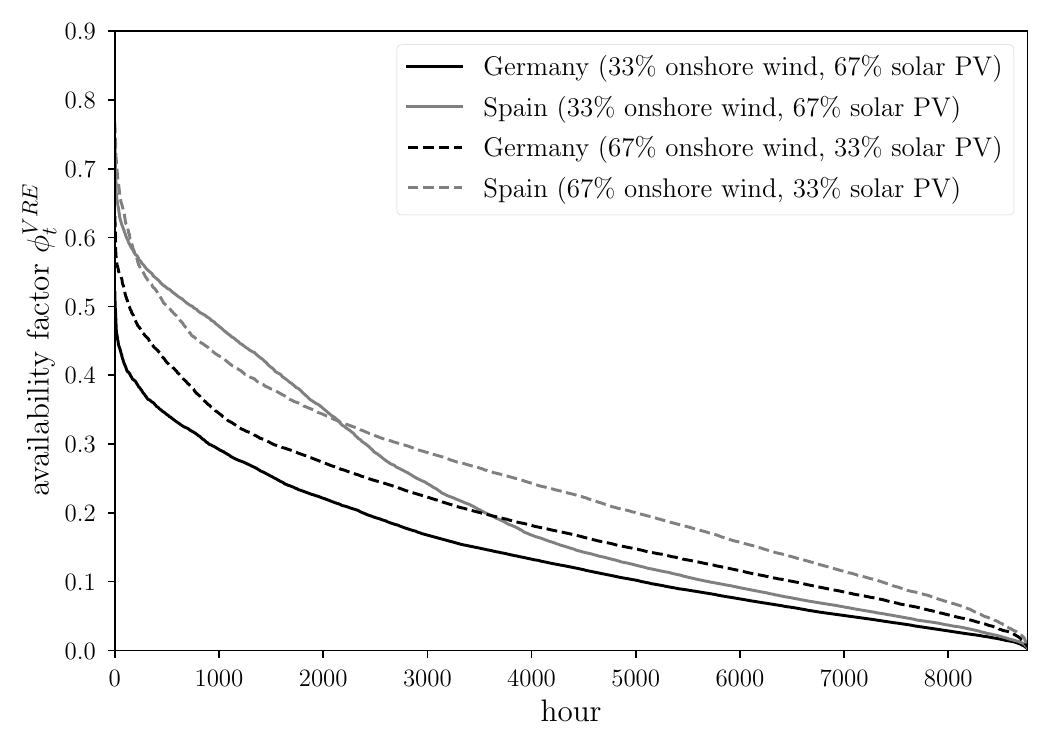}}}\\
\caption{Annual generation potentials of onshore wind power, solar \ac{PV}, and portfolios with different mixes of these technologies in Germany and Spain in 2010, visualized as the area below the availability duration curves. The data is retrieved from the Pan-European Climate Database \cite{de_felice_entso-e_2022}.}%
\label{fig:duration_curves}%
\end{figure}

Previous research draws on absolute, exogenously set thresholds for availability factors, e.g., 0.01, 0.02, 0.05, or 0.10 \cite{leahy_persistence_2013,patlakas_low_2017,kaspar_climatological_2019,ohlendorf_frequency_2020,kies_critical_2021,potisomporn_extreme_2024}. Yet, such analyses face compatibility issues when analyzing systems with different technologies and/or across several regions. This is because applying absolute thresholds on normalized availability factor time series implies assuming uniform generation capacity (e.g.,~in megawatts) across all compared systems. However, these systems may have very different annual generation potentials (e.g.,~in megawatt hours) depending on the meteorological conditions of the location and the \ac{VRE} technology. \ac{VRE} availability duration curves, sorting availability factor time series in descending order, highlight the discrepancies in generation potentials between different technologies and regions (Figure~\ref{fig:duration_curves_single}).

The same applies to energy systems transitioning towards variable renewables. Here, the optimal capacity mix of \ac{VRE} technologies depends on their respective generation potentials as well as cost parameters, and may vary substantially across regions. This results in diverse generation potentials for such \ac{VRE} portfolios based on the location and the capacity mix (Figure~\ref{fig:duration_curves_portfolio}).

\ac{FLH} serve as a commonly used metric for comparing annual average generation (or generation potential before potential curtailment). This metric indicates how long a generator (or technology aggregate) would operate at full capacity to provide its annual energy (potential). An analysis based on a uniform, exogenously set threshold for each compared system does not account for discrepancies in annual generation potentials. Consequently, such analyses (e.g., \cite{kaspar_climatological_2019,kies_critical_2021,ohba_climatology_2022,mayer_probabilistic_2023}) are biased by a \ac{FLH} effect, which puts their validity into question.

As a remedy, the drought threshold can be scaled to account for full-load hour differences between \ac{VRE} technologies and locations. Threshold scaling relative to different time series attributes can be found in the literature. One option is to scale thresholds relative to their system-specific maximum availability factor \cite{breyer_reflecting_2022}.\footnote{Technically, in this study, thresholds scale with the maximum generation. Discarding potential curtailment, this is equivalent to scaling with the maximum availability factor.}. Focusing on the \ac{VRE} availability of a single point in time (i.e.,~the peak point of the duration curves in Figure \ref{fig:duration_curves}), this approach ignores general availability patterns over the course of a year. Consequently, it does not mitigate differences in full-load hours.

Another option is to use a percentile of the availability time series distribution as threshold \cite{fleig_global_2006,cannon_using_2015,plain_accounting_2019}. Duration curves are a graphical representation of availability factor distributions (Figure~\ref{fig:duration_curves}). A percentile refers to the availability factor below a certain fraction of the population of time steps falls and depends on the shape of the distribution, which may vary across technologies and regions. Unlike scaling relative to the maximum availability factor, this method recognizes all \ac{VRE} availability states. However, the percentile approach does not fully adjust for \ac{FLH} differences.

An effective option for this is to align the drought threshold with each system's mean availability factor $\overline{avail}$. This method inherently scales with the system's full-load hours and, therefore, enables comparability across technologies or technology portfolios, regions, and, in the case of comparing multiple years, time \cite{raynaud_energy_2018,rinaldi_wind_2021,gangopadhyay_role_2022,otero_copula-based_2022,kapica_potential_2024}.

A priori, a meaningful parameterization of the fraction $frac$ of the mean availability factor $\overline{avail}$ for drought identification is not straightforward. Previous studies use various fractions of the mean \cite{yevyevich_objective_1967}. While lower thresholds, e.g., $thres = 0.1 * \overline{avail}$, find very severe but potentially brief and isolated \ac{VRE} drought events, higher fractions closer to $\overline{avail}$ may result in identifying extended periods of below-average availability, i.e., inter-annual variations in overall \ac{VRE} availability, rather than true \ac{VRE} droughts. Relevant fractions likely lie between these extremes. It therefore appears useful to explore a range of thresholds, $thres = frac * \overline{avail}$, for example with $frac$ ranging between 0.1 and 0.9. Obtained events reflect different degrees of low \ac{VRE} availability, which may be relevant for different types of flexibility options in renewable energy systems.

\subsection{Technological and spatial scope of analysis}\label{ssec:gen_scope}

The scope of a \ac{VRE} drought analysis focusing on \ac{VRE} supply may differ in terms of technological and spatial scope. Each of these dimensions demands specific assumptions, weighting, as well as comparability requirements.

In contrast to single-technology drought analyses, according investigations of a \ac{VRE} portfolio require a composite time series that combines the availability factors of different \ac{VRE} technologies. It can be computed as the weighted average of the technology-specific time series. Weighting factors correspond to the technology shares in the assumed overall capacity mix. Considering the temporal and spatial complementarity of wind and solar power \cite{weschenfelder_review_2020,pedruzzi_review_2023,lopez_prol_wind-solar_2024}, such a composite time series allows for within-portfolio balancing, which reflects the generation dynamics of power sectors with high shares of variable renewables. This is in line with several related studies \cite{raynaud_energy_2018,otero_characterizing_2022,ruhnau_storage_2022,francois_statistical_2022}. Another approach excludes or limits such balancing by applying a definition of compound \ac{VRE} drought events as periods with simultaneously arising drought conditions of multiple \ac{VRE} technologies \cite{rinaldi_wind_2021,mayer_probabilistic_2023}. This discards \ac{VRE} complementarity, hence potentially overestimating droughts characteristics of a \ac{VRE} portfolio.

For multi-regional settings, drought analyses using \ac{VRE} availability time series can be based on two extreme scenarios regarding regional interconnection\footnote{Here we abstract from within-region grid congestion.}: i) perfect interconnection, which abstracts from grid limitations (often referred to as the ``copper plate'' assumption), and ii) complete isolation, where regions are treated as ``island systems'' without electricity exchange. When comparing \ac{VRE} droughts of such island systems, \ac{FLH}-adjusted threshold scaling is required as discussed above. Realistic patterns of cross-regional electricity exchange fall between these extremes and typically necessitate more sophisticated energy system modeling beyond simple \ac{VRE} time series analysis.

Similar to the technology portfolio case, multi-regional settings with assumed perfect interconnection require a composite time series that collectively represents all regions. It can be generated as the weighted average of respective regional time series. The weights are assigned based on the relative share of each region's capacity within the overall system. Unless multiple cross-regional composite time series are to be compared, no \ac{FLH}-adjustment of the threshold is required. For \ac{VRE} portfolios across regions, weighting factors need to reflect both within-regional and across-regional shares in the capacity mixes.

\section{Extending the methodology to positive residual load events}\label{sec:prl}

\subsection{Positive residual load event definition}\label{ssec:prl_definition}

In addition to focusing exclusively on \ac{VRE} supply, research interest in the dynamics between \ac{VRE} droughts and periods of high electricity demand is intensifying. This becomes more relevant as many countries plan to electrify their heat, transport, and industry sectors, and respective additional loads enter the power sector. Notably during the heating season, electrified heating appliances such as heat pumps increase electricity demand with characteristic profiles \cite{bloess_power--heat_2018,ruhnau_heating_2020,roth_flexible_2023}. Compound high-demand and \ac{VRE} drought events, sometimes also referred to as ``cold dark doldrums'' or, in  German, ``kalte Dunkelflauten'' are emerging as a challenge for decarbonized energy systems that rely on \ac{VRE}.

\ac{PRL} events, characterized by high demand and a shortage of \ac{VRE} supply, indicate the need for firm capacity from thermal generators, or flexibility provided by electricity storage, imports, or demand-side measures \cite{schill_electricity_2020,zerrahn2018,goke_how_2023}. In contrast, negative residual load events represent \ac{VRE} surplus generation. This surplus can be integrated into the power sector through electricity storage or exports to other areas, or used for electrified loads from coupled sectors, such as heat provision or transport \cite{brown_synergies_2018}. Alternatively, it can be curtailed \cite{schill_residual_2014, zerrahn2018}. Note that the latter is not an option for \ac{PRL} periods, which makes them more challenging to deal with than negative residual loads. Accurately quantifying both positive and negative residual load events is relevant for capacity planning of weather-resilient energy systems.

\begin{figure}[htbp]
\centering
\noindent\includegraphics[width=0.97\linewidth,height=\textheight, keepaspectratio]{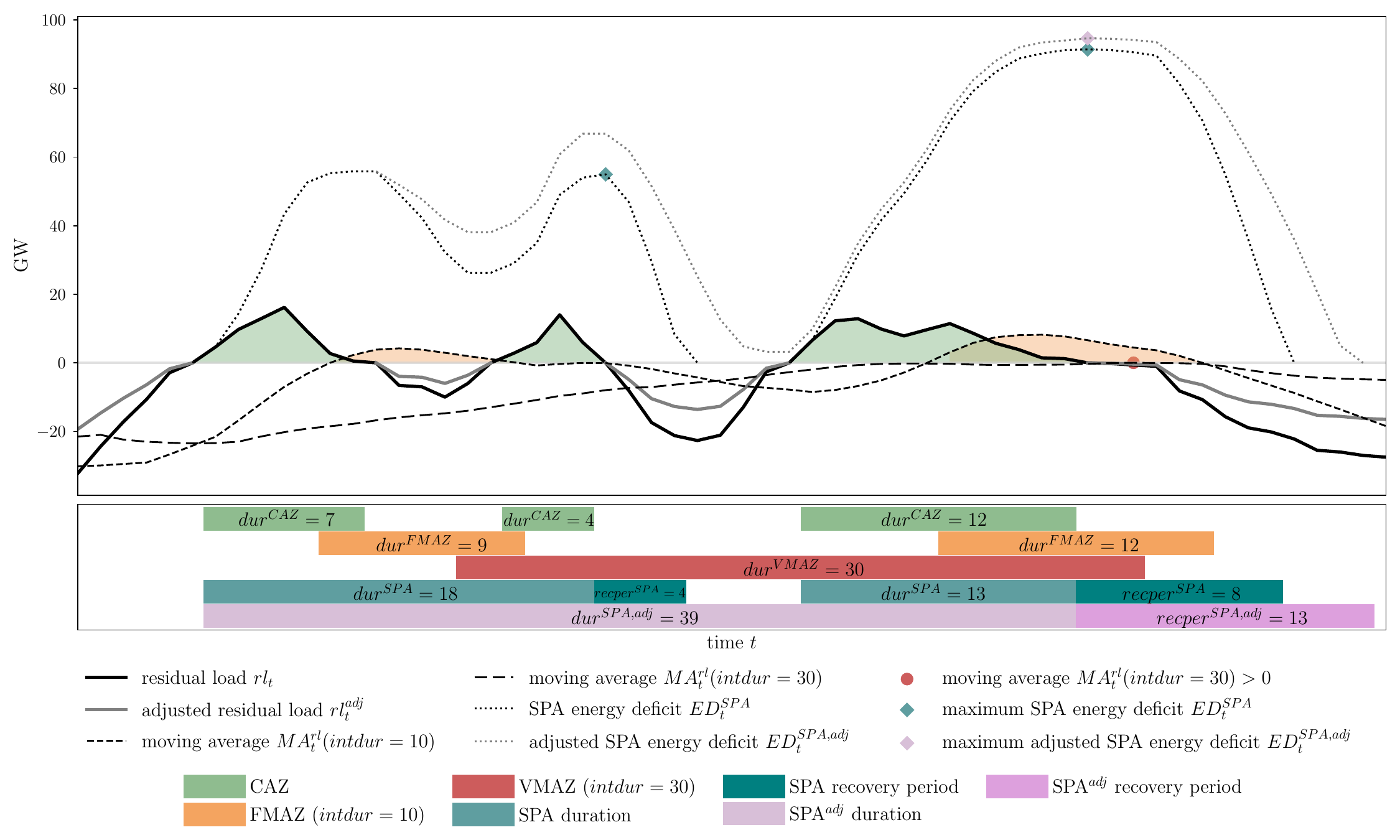}
\caption{Exemplary residual load time series and identification methods \ac{CAZ}, \ac{FMAZ}, \ac{VMAZ}, \acf{SPA} (with or without recovery), and \ac{SPA$^{adj}$} (upper panel). The green and orange areas indicate the obtained \ac{CAZ} and \ac{FMAZ} energy deficit, respectively. The lower panel's horizontal bars illustrate the varying event duration $dur$ identified by each method.}
\label{fig:rl_cbt_mbt_spc}
\end{figure}

Key \ac{PRL} characteristics, e.g.,~duration and energy deficits, can be analyzed using residual load time series $rl_t$ (Figure~\ref{fig:rl_cbt_mbt_spc}). While this can be done analogously to \ac{VRE} drought analysis, the energy deficit of \ac{PRL} events can directly be associated with the need for energy that has to be provided by, for instance, long-duration storage. Importantly, the maximum capacity need can be inferred from peak residual load and does not require a dedicated analysis of \ac{PRL} events.

Unlike \ac{VRE} drought analysis, there is no need for deliberate threshold parameterization to identify \ac{PRL} events. Instead, \ac{PRL} events are simply obtained as periods above zero (gray line in Figure~\ref{fig:rl_cbt_mbt_spc}). In previous \ac{PRL}, analysis other approaches such as percentile-based thresholds were used \cite{otero_copula-based_2022}. However, positive residual load directly relates to the need for energy from firm generators, storage, or imports. Hence, from the perspective of a power sector, diverting from thresholds aligned with the zero line appears counter-intuitive.

\subsection{Positive residual load event identification methods}\label{ssec:vre_window_event_cbt_mbt_spc}

We adapt the \ac{VRE} droughts identification methods discussed in Section~\ref{sec:vre_droughts_supply} to \ac{PRL} events. Unless stated otherwise, the algorithmic routine is equivalent but recalibrated to identify periods where residual load $rl_t$, or its average $MA_t^{rl}$, is above zero, rather than below a designated threshold. Analogously to \ac{CBT} and \ac{MBT}, we denote \acf{CAZ} and \ac{MAZ} as:

\begin{align}
    PRL_t^{CAZ} = &
    \begin{cases}
        1 & \text{if $rl_t > 0 \ \ \forall \ \ t$}\\
        0 & \text{if $rl_t \leq 0 \ \ \forall \ \ t$}
    \end{cases}
    &
    PRL_t^{FMAZ} = &
    \begin{cases}
        1 & \text{if $MA_t^{rl}(intdur) > 0 \ \ \forall \ \ t$}\\
        0 & \text{if $MA_t^{rl}(intdur) \leq 0 \ \ \forall \ \ t$}
    \end{cases}
\end{align}

Accordingly, the \acf{FMAZ} and \acf{VMAZ} concepts emulate the \ac{FMBT} and \ac{VMBT} approaches.

Analogously to \ac{VRE} drought analysis, the \ac{PRL} event duration $dur^{CAZ}$ and $dur^{FMAZ}$ corresponds to the number of consecutive time steps with $PRL_t^{CAZ}=1$ and $PRL_t^{FMAZ}=1$, respectively. The duration of \ac{VMAZ} event equals the iteration-specific averaging interval duration $dur^{VMAZ} = intdur^{VMAZ}$.

The \ac{VMAZ} energy deficit is near zero and hence insignificant. With $t_k$ and $t_l$ as the start and end points of a \ac{PRL} event, the \ac{CAZ} and \ac{FMAZ} event's energy deficit (green and orange areas in Figure \ref{fig:rl_cbt_mbt_spc}) are defined as: 

\begin{align}
    ED_t^{CAZ} = & \sum_{t_k}^{t_l} rl_t 
    & 
    ED_t^{FMAZ} = & \sum_{t_k}^{t_l} MA_t^{rl} & \forall \ \ t_k \leq t \leq t_l
\end{align}

The limitations associated with \ac{CBT}, \ac{FMBT}, and \ac{VMBT} discussed in Section~\ref{ssec:vre_window_event_cbt_mbt_spc} also extend to their \ac{PRL} counterparts. The \ac{CAZ} method, while directly linking event duration, its energy deficit, and its temporal occurrence to for example the required generation and dispatch period of a capacity-constrained firm generator, ignores brief periods of negative residual load adjacent to \ac{PRL} events (compare the two \ac{CAZ} events on the left in Figure~\ref{fig:rl_cbt_mbt_spc}). This can lead to underestimating the need for energy-constrained flexibility options, such as long-duration electricity storage. As for \ac{FMAZ}, the parameterization of the averaging interval duration $intdur^{MAZ}$ lacks a directly attributable power sector reference and remains arbitrary. As identified \ac{FMAZ} events may substantially change with varying $intdur^{MAZ}$, they are ambiguous. Further, their interpretation is not straightforward. In general, longer averaging periods may lead to an underestimation of actual \ac{PRL} challenges as averaging already assumes temporal balancing, for instance, in real-world power systems carried out by electricity storage or other flexibility options. The \ac{VMAZ} approach, in contrast, obtains unique events. Yet, its relevance remains limited in \ac{PRL} analysis as it only allows for evaluating \ac{PRL} event duration but not energy deficit, which converges to zero by design.

The \ac{SPA} identifies unique \ac{PRL} while obtaining a meaningful energy deficit. Unlike for \ac{VRE} droughts, \ac{SPA} does not require setting an arbitrary threshold for \ac{PRL} detection, precluding a respective shortcoming in the case of \ac{VRE} drought identification. \ac{SPA} allows for brief periods with negative residual load and detects all \ac{PRL} events within a single iteration. Consequently, we regard it as particularly well-suited for identifying \ac{PRL} events. Analogously to \ac{VRE} droughts, its cumulative energy deficit is denoted as:

\begin{equation}
    ED_t^{SPA} = 
    \begin{cases}
        ED^{SPA}_{t-1} + rl_t & \text{if $ED^{SPA}_{t-1} + rl_t > 0 \ \ \forall \ \ t$}\\
        0 & \text{if $ED^{SPA}_{t-1} + rl_t \leq 0 \ \ \forall \ \ t$}
    \end{cases}
\end{equation}

The \ac{SPA} event duration $dur^{SPA}$ refers to the period beginning when the cumulative energy deficit of a single SPA event becomes positive to its global maximum. The \ac{SPA} method further allows for quantifying a recovery period $recper^{SPA}$. Originating from the hydrology field \cite{duckstein_engineering_1987}, this period represents the balancing time required for the surplus \ac{VRE} energy from intermediate and subsequent negative residual load events to offset the energy deficit of a \ac{PRL} event. It corresponds to the period following the peak cumulative energy deficit $ED_t^{SPA}$ to the point in time where this deficit returns to zero $ED_t^{SPA} = 0$ (Figure~\ref{fig:rl_cbt_mbt_spc}).

In real-world energy systems, long-duration electricity storage is a key option for dealing with \ac{PRL} events \cite{ruhnau_storage_2022}. Storage always incurs conversion losses, i.e.,~the amount of electricity recovered from storage is always smaller than previous storage charging. The standard \ac{SPA} method does not account for such losses. To address this, we introduce an \acf{SPA$^{adj}$} method, which incorporates storage efficiency losses arising when balancing positive and negative residual load periods.\footnote{For simplicity, we assume that electricity storage does not have any binding capacity constraints with respect to charging, discharging and storage energy.} This can be achieved by adjusting all negative residual load values to account for storage round-trip\footnote{Technically, this adjustment allots the conversion losses from both charging and discharging to storage charging occurring in intermediate and subsequent negative residual load periods, whose surplus energy is used to offset the energy deficit of a \ac{PRL} event.} efficiency $eff^{LDS}$ as denoted by:

\begin{equation}
    rl_t^{adj} = \frac{rl_t}{eff^{LDS}} \ \ \forall \ \ rl_t < 0
\end{equation}

In the \ac{SPA$^{adj}$} method, the adjusted energy deficit $ED_t^{SPA,adj}$ is calculated equivalently to $ED_t^{SPA}$. However, assuming that long-duration electricity storage is the relevant flexibility option for dealing with a \ac{PRL} event, the round-trip efficiency adjustment prolongs the recovery period $recper^{SPA,adj}$ to reflect the time needed to recharge the storage to its pre-\ac{PRL} state of charge (Figure~\ref{fig:rl_cbt_mbt_spc}). This approach results in an extended recovery period $recper^{SPA,adj}$ and event duration $dur^{SPA}$ when compared to calculations without considering efficiency losses. This adjustment leads to a more realistic representation of the energy system flexibility need during compound peak demand and \ac{VRE} drought events. Note that real-world storage options may also incur relevant standing losses, depending on how long the energy is stored. Incorporating these into the \ac{SPA} concept would be significantly increase complexity and should rather be studied using appropriate energy models.

\subsection{Comparability, scope of analysis, and critical appraisal}

Unlike \ac{VRE} droughts, analyzing \ac{PRL} events does not require pre-processing, such as a \ac{FLH} adjustment, to enable comparability. \ac{PRL} events are denoted in energy (e.g., megawatt hours), which is readily comparable across different technological, spatial, and temporal scopes.

The analysis of residual load time series can be approached in two ways. One option is to use exogenous assumptions for both \ac{VRE} availability factors and demand time series. This setup facilitates the exploration of various user-specific \ac{VRE} capacity mix scenarios.

Alternatively, residual load time series can be retrieved from a scenario optimized by an energy system model. This method is more resource-intensive due to the need for comprehensive input data and complex modeling assumptions, but also less dependent on exogenous capacity assumptions. It further enables an analysis of policy-relevant scenarios, such as those outlined in strategic documents like the Ten Year Network Development Plan for the European electricity infrastructure \cite{entso-e_ten_2022}.

Characterizing extreme \ac{PRL} events, i.e., their duration, energy deficit, frequency, etc., allows exploring the need for power sector flexibility, which can be provided by long-duration storage or other flexibility options. The \ac{SPA$^{adj}$}, accounting for energy losses of such flexibility options, appears to be particularly useful in this respect.

During an extreme \ac{PRL} events, brief periods with very high \ac{VRE} generation requiring partial curtailment may occur. None of the discussed identification methods is able to account for such curtailed energy, which can lead to underestimating the energy deficit and duration of extreme \ac{PRL} events \cite{ruhnau_storage_2022}. More importantly, insights from \ac{PRL} event analysis are only applicable to specific assumptions on the \ac{VRE} portfolio and load, or driven by the parameterization of optimizing models. Therefore, when discussing characteristics such as the duration, frequency, or energy deficit of \ac{PRL} events, contextualization is essential. This is particularly important for a meaningful comparison of the findings of different studies and for communication with policy-makers or the general public.

By design, an exogenously derived residual load time series does not account for electricity exchange and potentially neglects flexibility from conventional demand and electrified loads from the heating, transport, or industry sector. Accurately estimating the operation of these loads without the use of an energy system model is challenging and may require numerous assumptions.

As for residual load time series generated by optimizing energy system models, model outcomes rely on the used modeling framework and its underlying optimization rationale, the level and detail of technology representation, and techno-economical assumptions, such as available technologies, costs, meteorological resources \cite{gils_modeling_2022,gils_model-related_2022,van_ouwerkerk_impacts_2022}. For instance, ceteris paribus, changes in storage investment costs can significantly affect storage investment and usage in a modeled scenario, thus altering the energy deficit or duration of obtained \ac{PRL} events. Further, an optimized residual load already takes into account the endogenously determined investment and dispatch decisions of electricity storage and other flexibility options. This inherently affects \ac{PRL} event characteristics.

\section{Conclusion and recommendations}\label{sec:conclusion}

\subsection{Summary}

\ac{VRE} shortage has drawn research interest from various fields, including hydrology, wind and solar energy, energy system modeling, or climatology, and has been analyzed using a range of methods and disparate naming. Yet, there is no structured overview and explanation of these methods using harmonized terminology. It thus remains unclear as to what methodological approach is useful for different settings of \ac{VRE} shortage analysis, which can be studied based on time series of \ac{VRE} availability or positive residual load. For clarity, we suggest ``variable renewable energy shortage'' as a generic term  covering both ``variable renewable droughts'' as periods with low resource availability of a single or multiple \ac{VRE} technologies and ``positive residual load events'' as periods with a net imbalance of electric load and \ac{VRE} supply.

This paper addresses this gap by providing a methodological overview and guidance for defining and quantifying \ac{VRE} shortage events across technologies and regions, discussing different approaches for both \ac{VRE} availability and \ac{PRL} time series. While not claiming to be exhaustive, we categorize and discuss relevant existing shortage event definitions and identification methods, propose methodological improvements, and elaborate on how to enable compatibility in cross-technology or multi-regional analyses. Our work seeks to support the harmonization of terminology across related fields. We also aim to aid in advancing future research on \ac{VRE} shortage events by facilitating a better interpretation of such studies' potential findings.

\subsection{What (not) to do in variable renewable energy shortage analysis}

We identify a range of good practices for \ac{VRE} shortage analysis to obtain accurate, broadly applicable, and easy-to-interpret findings on the duration, frequency, and energy deficit of \ac{VRE} shortage events. This is important for guiding policy or infrastructure decisions for future energy systems with high \ac{VRE} penetration.

We advocate for future work to clearly and unambiguously specify the shortage period definitions and identification methods used, as well as key parameter choices such as thresholds or averaging intervals. Methods that are poorly defined or not transparent hinder the accurate interpretation and contextualization of findings, in both research and policy discourse.

For accurate identification, \ac{VRE} shortage analysis should meet a range of methodological criteria. First, \ac{VRE} shortage periods should be defined according to the event perspective, i.e., capturing the full extent of qualifying periods allowing for event duration ranging from just a few time steps to extensive weeks or even months. Second, identified shortage events should be unique, avoiding double counting and overlap with adjacent events. Third, the method should allow for pooling of closely occurring events to include periods that independently may not qualify as \ac{VRE} drought or \ac{PRL} events but are adjacent to periods of low availability or high residual load.

In general, reporting the \ac{VRE} shortage duration appears to be equally relevant to \ac{VRE} drought and \ac{PRL} analyses. Yet, claims on identified durations have to be contextualized, as they depend on the identification method used. Further, evaluating energy deficits of \ac{PRL} events is valuable for understanding the demand for firm generation or long-duration storage energy in renewable power sectors. Conversely, even though this metric is commonly applied in \ac{VRE} drought analysis across different bodies of literature, it appears to be less relevant in such settings, as it strongly depends on the threshold parameter choice and cannot be associated with a power sector attribute.

For \ac{VRE} drought analysis, the \ac{VMBT} and \ac{SPA} methods align well with these methodological requirements. The \ac{VMBT} method finds events with the longest duration but lacks meaningful energy deficits, which is a reasonable cost for accurate duration identification. In the case of \ac{PRL} analyses, we recommend the use of \ac{SPA} or its adjusted version, \ac{SPA$^{adj}$}. Besides adhering to the outlined criteria, these methods allow for accurately quantifying both duration and the energy deficit of \ac{PRL} events.

Analyzing \ac{VRE} droughts across multiple regions and/or for different renewable technologies or technology portfolios poses comparability challenges. It is essential to base the analysis on uniform average generation (or generation potential) across compared systems to avoid a full-load hour bias in obtained drought characteristics. Threshold scaling for each investigated system according to a fraction of the mean availability factors is an effective option for enabling comparability. 

The selection of the respective fraction remains a sensitive parameter choice and requires careful consideration. Too high thresholds may identify long-lasting periods with below-average renewableavailability rather than actual \ac{VRE} droughts. Conversely, very low thresholds may find very severe but potentially brief and isolated droughts. We advise exploring various threshold values between these extremes, ranging from near-zero to the average availability factor. This ensures a comprehensive capture of the full spectrum of drought events, from mild to very severe.

For transparency and reproducibility, it further appears vital to document and provide the underlying data and the algorithms used, adhering to open-source and open-data principles \cite{pfenninger_opening_2018}. Such practices will not only foster transparency and replicability but also strengthen the overall accuracy and practical usefulness of \ac{VRE} drought analyses.

Based on our analysis, we identify a range of bad practices for variable renewable energy shortage analysis. Searching for fixed-duration shortage windows underestimates shortage duration while inflating the number of identified shortage periods. We thus strongly recommend not using the window perspective. Methods yielding ambiguous sets of events that may change for varying method parameterization, such as \ac{FMBT} and \ac{FMAZ}, should be avoided due to the limited applicability of obtained shortage characteristics. Next, we advise against the use of methods incapable of pooling mutually dependent shortage events, such as \ac{CBT} and \ac{CAZ}, as obtained results overestimate the number of events while underestimating event duration and energy deficit. Finally, in \ac{VRE} drought studies, using absolute thresholds can further lead to misleading comparisons based on normalized generation capacities rather than actual energy production, resulting in potentially flawed conclusions.

\subsection{Methodological trade-offs and outlook}

Renewable energy shortage can be studied based on time series of \ac{VRE} availability or positive residual load. While both approaches have their merits, a trade-off between policy relevance and the universality of insights emerges. Drought analyses based on \ac{VRE} availability data typically require fewer assumptions, yielding conclusions with broader applicability. \ac{PRL} events analyses aim to be more policy-oriented, as they quantify the mismatch between electricity demand and variable renewable supply, which is especially relevant for highly renewable decarbonization scenarios. However, the latter approach requires extensive assumptions on the composition of the renewable generation portfolio and demand-side flexibility. This may necessitate detailed energy system modeling, whose outcomes hinge upon underlying optimization rationales, technology representation, and parameterization. This potentially limits the validity of \ac{PRL} analyses to the particular scenarios investigated.

Such a trade-off also applies to the technological scope of \ac{VRE} drought analysis. Aside from methodological assumptions inherent to the generation of variable renewable availability time series, e.g., such for reanalysis, insights from single-technology analysis are broadly applicable but less policy-relevant for future energy systems that rely on a mix of renewable technologies. Conversely, analyzing a \ac{VRE} portfolio is more policy-oriented but involves energy modeling or exogenous assumptions on the composition of the investigated capacity mix. \ac{VRE} capacity mixes determined by an energy model are optimal only in the context of the other optimized system components and flexibility options that are represented in the model (e.g., storage, electricity exchange, or direct-electrified applications from the heat, industry, or transport sector). Similarly, exogenous capacity mixes rely on user-specific assumptions. These aspects limit the validity of insights to investigated scenarios.

While we provide general guidance on the advantages and drawbacks of different \ac{VRE} shortage identification methods, a thorough quantitative comparison would be desirable, ideally for a range of variable renewable technologies and electric load profiles for different world regions and weather years. In the Supplementary Information, we briefly indicate how a comprehensive comparative study could be carried out based on a stylized case study. Similar analyses have been conducted by Ohlendorf \& Schill \cite{ohlendorf_frequency_2020} and Potisomporn et al. \cite{potisomporn_extreme_2024} for a subset of methods and technologies as well as a limited number of regions. Another key aspect that requires attention is the choice of the threshold parameter, particularly in distinguishing between severe \ac{VRE} droughts and prolonged periods of below-average availability. This distinction could support the selection of relevant weather years to enable weather- and climate-resilient energy system analyses and well-informed decarbonization policies. Additionally, assessing the sensitivity of outcomes to different \ac{VRE} availability datasets would be useful and could be valuable feedback to the meteorological research community that provides and continuously improves such datasets \cite{craig_overcoming_2022}.

\section*{Acknowledgments}
We thank the entire research group ``Transformation of the Energy Economy'' at the German Institute for Economic Research (DIW Berlin) for valuable inputs and discussions, as well as conference participants of the EGU General Assembly 2022, the International Energy Workshop 2022, the Conference on Climate, Weather and Carbon Risk in Energy and Finance 2022, the Next Generation Energy and Climate Workshop 2022, and the International Conference Energy \& Meteorology 2023 for valuable comments on earlier drafts. We acknowledge research grants by the Einstein Foundation (grant no.~A-2020-612) and by the German Federal Ministry of Education and Research via the ``Ariadne'' projects (Fkz 03SFK5NO \& 03SFK5NO-2).

\section*{Author Contributions}\label{sec:author contributions}


\textbf{Martin Kittel}: Conceptualization (equal), methodology, software, investigation, data curation, visualization, writing - original draft, review and editing (equal). \textbf{Wolf-Peter Schill}: Conceptualization (equal), writing - review and editing (equal), project administration.

\bibliography{references}

\newpage

\renewcommand{\thesection}{SI}
\global\long\def\thefigure{SI.\arabic{figure}}
\setcounter{figure}{0}

\global\long\def\thetable{SI.\arabic{table}}
\setcounter{table}{0}

\global\long\def\theequation{SI.\arabic{equation}}
\setcounter{equation}{0}

\setcounter{page}{1}

\section{Supplementary information}\label{sec:sup_inf}

\subsection{Stylized example for a comparative study of shortage period definitions and identification methods}

In addition to the methodological overview provided in the main part of this paper, a comprehensive quantitative comparison of all elaborated approaches would be of interest to quantify our general recommendations on how (not) to analyze \ac{VRE} shortage periods. Such a comparison could examine the effects of using different shortage period definitions (e.g., window vs. event perspective) as well as identification methods for identifying \ac{VRE} droughts (e.g.,~\ac{CBT}, \ac{FMBT} for a range of averaging intervals $\omega$, \ac{VMBT}, \ac{SPA}, and \ac{SPA$^{adj}$}) or \ac{PRL} events (e.g., \ac{CAZ}, \ac{FMAZ} for a range of averaging intervals $\omega$, \ac{VMAZ}, \ac{SPA}, and \ac{SPA$^{adj}$}). 

Variable renewable energy sources differ significantly in their generation patterns and seasonality depending on the latitude of their location. A comparative study would therefore have to explore the technological (e.g., for on- and offshore wind, solar \ac{PV}, or hydro inflows) and regional impacts on renewable shortage identification, ideally for different regions on a country or maybe even sub-country level. It could also be of interest to incorporate a wide range of drought thresholds scaled relative to the respective mean of the time series. Finally, to enhance the validity of results, such a comparison would ideally cover a large number of years. Possible indicators for the analysis include frequency-duration distributions, return periods, maximum durations, seasonality, spatio-temporal correlation, and in the case of \ac{PRL} events, also energy deficits.

Yet, a comprehensive comparative study as outlined above is beyond the scope of this paper. However, we illustrate the potential for such a comparison with a stylized case study of onshore wind in Germany. Using the \ac{CBT} and \ac{VMBT} methods, we identify wind drought events over a six-year period (1990-1995) \cite{de_felice_entso-e_2022}. Given that this analysis focuses on a single technology and region, scaling to account for differences in technologies and regions, as outlined in section~\ref{ssec:vre_scaling}, is not required. For simplicity, we apply an absolute threshold of 0.1, which has often been used in the literature before.

\begin{figure}[htbp]
\centering
\noindent\includegraphics[width=0.97\linewidth,height=\textheight, keepaspectratio]{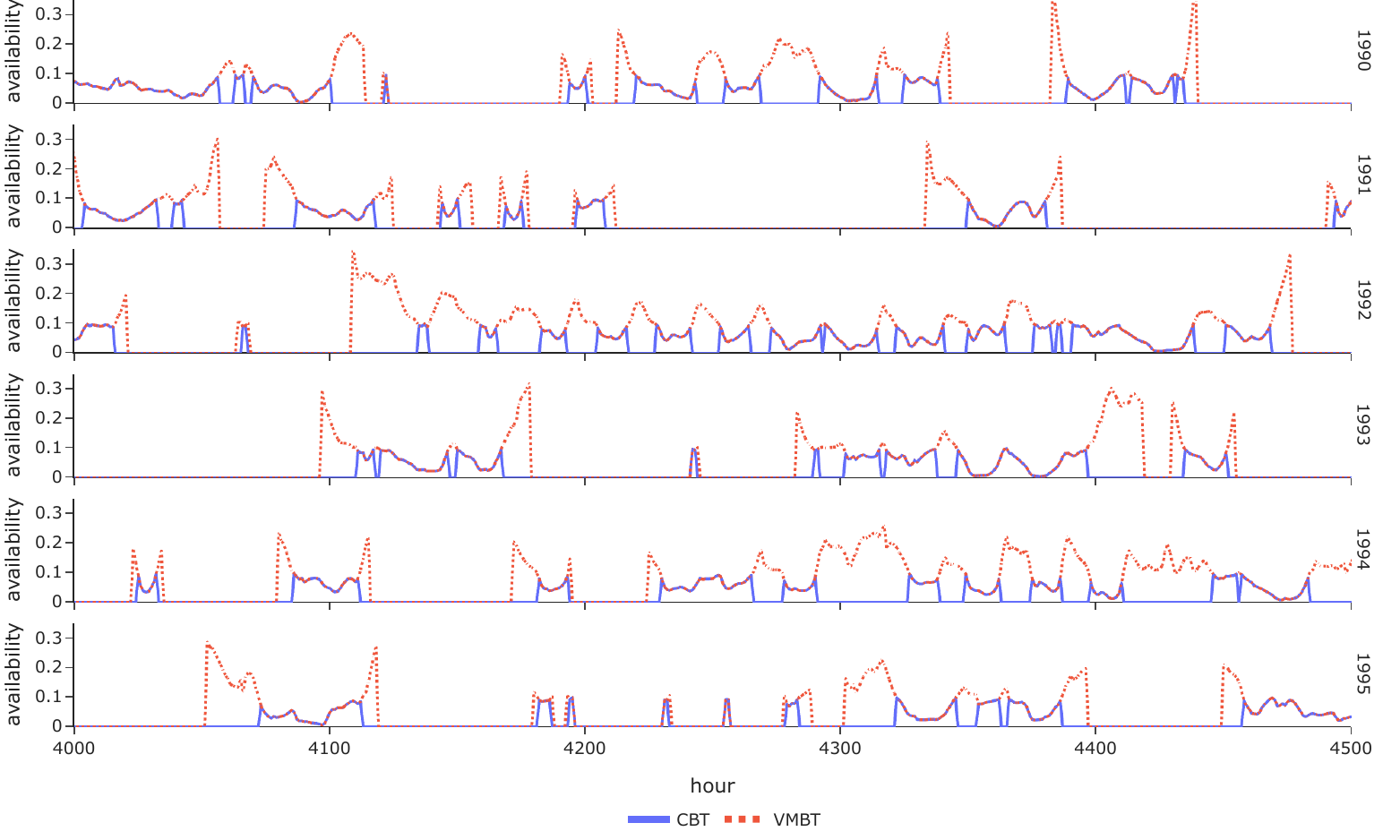}
\caption{Drought events identified with two different methods \ac{CBT} and \ac{VMBT} for exemplary summer periods. White space indicates periods that do not qualify as drought.}
\label{fig:case_study_ts}
\end{figure}

\begin{figure}[htbp]
\centering
\noindent\includegraphics[width=0.97\linewidth,height=\textheight, keepaspectratio]{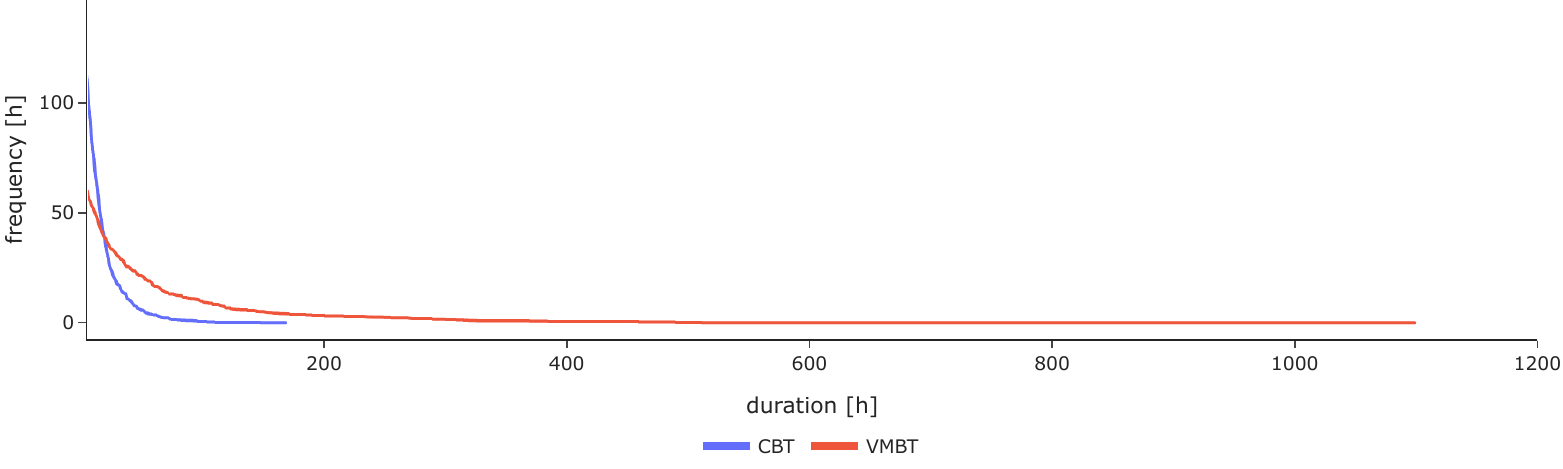}
\caption{Cumulative frequency-duration distribution of wind drought events identified by the \ac{CBT} and \ac{VMBT} methods, aggregated for the years 1990-1995.}
\label{fig:case_study_frequency}
\end{figure}

Figure~\ref{fig:case_study_ts} illustrates identified wind droughts over a 500-hour period in summer. The \ac{CBT} approach detects numerous brief periods constantly below the 0.1 threshold, whereas the \ac{VMBT} method identifies fewer but longer-lasting events. Multiple brief \ac{CBT} events occurring in close proximity to each other suggest mutually dependent events, which combined have a mean availability near or below the drought threshold. Methods such as \ac{VMBT} facilitate the pooling of such events, which is particularly visible in 1992. The frequency-duration distributions confirm these findings across all investigated years (Figure~\ref{fig:case_study_frequency}).

\begin{figure}[htbp]
\centering
\noindent\includegraphics[width=0.97\linewidth,height=\textheight, keepaspectratio]{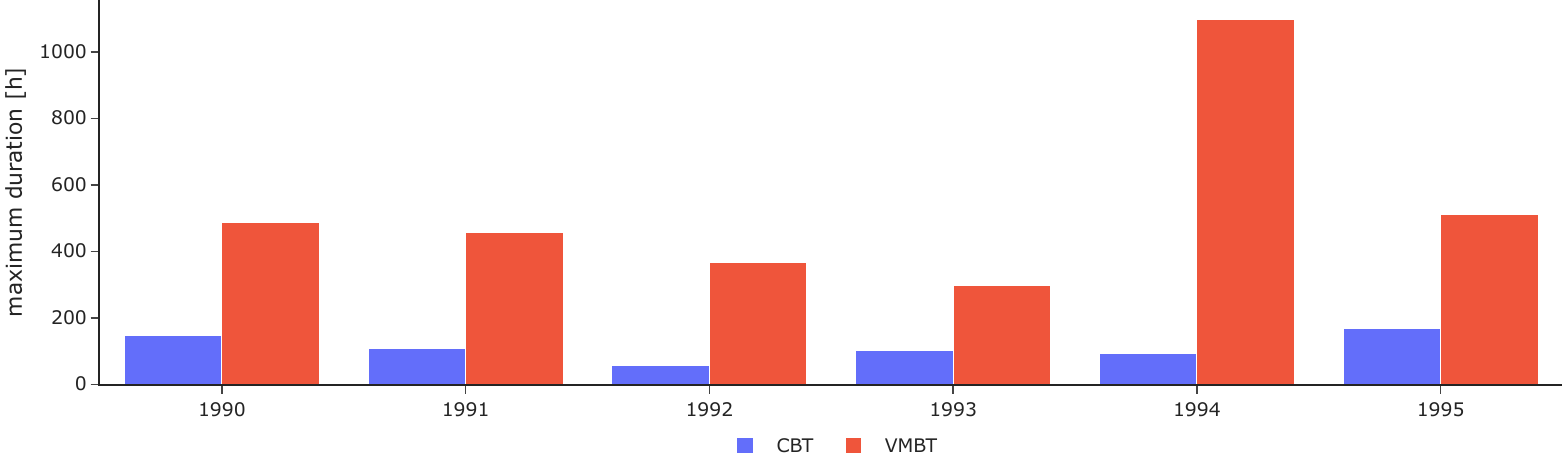}
\caption{Exemplary maximum drought duration identified using two identification methods \ac{CBT} and \ac{VMBT} based on the drought event notion.}
\label{fig:case_study_max_yearly}
\end{figure}

Our stylized example further shows that the duration of the most extreme droughts can vary significantly, depending on the weather year and the identification method used (Figure~\ref{fig:case_study_max_yearly}). In our setting, the longest \ac{CBT} drought lasts 168 hours and occurred in 1995. In contrast, the \ac{VMBT} method identifies a more than six times longer maximum drought duration of 1098 hours of a wind drought arising in 1994.

In future energy systems that heavily rely on variable renewable energy sources, extreme droughts necessitate substantial temporal and spatial system flexibility provided by storage, demand response, or geographical balancing via transmission. The need for system flexibility can be assessed with energy system models that use renewable availability time series as input. Due to computational limits, often only one or a few years can be used in such models. However, for weather-resilient energy system modeling, it is imperative to consider input data that ideally includes the most extreme droughts. Investigating renewable availability time series with the methods described above allows deliberately selecting relevant weather years, which can then be investigated with numerically constrained energy system models. 

In this respect, our stylized example indicates that \ac{CBT} results tend to underestimate the maximum duration of droughts that are relevant for renewable energy systems, and tend to overestimate the number of drought events. In contrast, the \ac{VMBT} method provides a wind drought characterization that is more relevant for energy system analysis. This facilitates a deliberate selection of relevant weather years, which serves as key input data for renewable energy system modeling. 

In future work, more detailed quantitative comparisons of the different methods for renewable energy shortage analysis that are presented in the main part of our manuscript could be carried out. This could raise insights on how shortage identification results differ between methods when different renewable technologies, weather years, and regions for various thresholds are considered. The stylized example presented above may inspire such future work. As a next step, systematically quantifying the impact of weather year selection on energy system outcomes would be desirable to evaluate the downstream effects of different renewable shortage period definitions and identification methods and could also be included in more comprehensive future comparative studies.

\end{document}